\documentclass[aps,prb,superscriptaddress,floatfix,10pt,twocolumn]{revtex4-2}
\usepackage{graphicx,graphics}
\usepackage{dcolumn}
\usepackage{amsmath,amssymb,amsfonts}
\usepackage{latexsym,verbatim}
\usepackage{bm}
\usepackage{color}
\usepackage{cancel}
\usepackage{ulem}
\usepackage{multirow}
\usepackage[abs]{overpic}
\usepackage[breaklinks=true,colorlinks,citecolor=blue,linkcolor=blue,urlcolor=blue]{hyperref}
\usepackage{array}
\usepackage{nicefrac}
\usepackage{pict2e}
\usepackage{float}
\usepackage{physics}
\usepackage{bbold}
\usepackage{soul}
\usepackage{enumerate}
\usepackage{hyperref}
\usepackage{xcolor}
\usepackage{upgreek} 
\usepackage{braket}

\begin{document}

\title{High-performance Andreev interferometer-based electronic coolers}

\author{Francesco Cioni}
\affiliation{NEST, Scuola Normale Superiore, Piazza San Silvestro 12, I-56127 Pisa, Italy}
\author{Fabio Taddei}
\affiliation{NEST, Istituto Nanoscienze-CNR and Scuola Normale Superiore, Piazza San Silvestro 12, I-56127 Pisa, Italy}

\begin{abstract}
In this paper, we propose an electronic refrigerator based on a ballistic Andreev interferometer that allows to reach a maximum cooling power per channel up to five orders of magnitude larger than that of the conventional normal metal–insulator–superconductor cooler.
This effect is achieved by exploiting
the destructive interference that occurs when the superconducting phase difference equals $\pi$. This results in a strongly suppressed charge current below the superconducting gap, while still allowing the extraction of excitations above the gap, leading to a cooler with enhanced performance. Interestingly, we find that such a large cooling power per channel enables the achievement of an electronic temperature close to the theoretical lower bound. Additionally, we derive an approximate expression for this bound in the regime of low bath temperatures. Finally, we propose potential implementations of the ballistic Andreev interferometer cooler using semiconductors, graphene, and topological insulators.
\end{abstract}

\maketitle

\section{Introduction}
The importance of heat management in nanoscale devices and quantum technologies~\cite{Giazotto2006,Benenti2017,Pekola2021,Majidi2024} is becoming increasingly crucial essentially because of the need to maintain lower temperatures.
Quantum devices, on the other hand, offer new possibilities for the control of heat fluxes, such as 
the exploitation of interference effects (for example through the Aharonov-Bohm phase).
Such effects have been considered in several recent papers
to realize thermal machines~\cite{Haack2019,Haack2021}, to manage heat fluxes ~\cite{Giazotto2012,Martinez2013,Bosisio2015,Fornieri2016,Solinas2016,Fornieri2017,Sothmann2017,Hwang2018,Acciai2021,Huang2023,Hwang2024,Balduque2024} and cooling~\cite{Fornieri2016,Hwang2024} or for thermoelectric effects even in hybrid superconducting systems~\cite{Blasi2023}.

In nanoscale quantum systems, excessive heat buildup can lead to decoherence and performance degradation. Various methods for achieving electronic refrigeration have been proposed and implemented to address this issue. For a comprehensive review of these techniques, see Refs.~\cite{Giazotto2006,Muhonen2012,Courtois2014,Ziabari2016,Jones2020,Cao2021}.
The possibility of using normal metal–insulator–superconductor (NIS) junctions for electron cooling was first proposed in the 1990s in Refs.~\cite{Nahum1994,Bardas1995,Leivo1996}, and thereafter studied in many papers~\cite{Fisher1999,Savin2001,Pekola2004,Clark2005,Vasenko2010,Quaranta2011,Rajauria2012,Lowell2013,Nguyen2014,Camarasa2014,Nguyen2015,Zhang2015,Gunnarsson2015,Nguyen2016,Kashiwaya2016,Courtois2016,Sanchez2017,Sanchez2018,Marchegiani2018,Kapri2018,Kapri2019,Hussein2019,Gordeeva2020,Vischi2020,Kemppinen2021,Hwang2023,Hatinen2024,Verma2024}.
Such a scheme relies on the fact that quasiparticle transfer from N to S is possible only at energies greater than the superconducting gap $\Delta$ in such junctions. By applying a voltage bias, electrons can be thus removed from the "hot tail" of the electronic distribution, effectively reducing the electronic temperature in N. However, this cooling mechanism is limited by the possibility of sub-gap tunneling of Cooper pairs, which is accounted for by Andreev reflection [consisting in the backscattering of an electron into a hole at the normal-superconductor (NS) interface].
Indeed, these processes do not transfer energy from N to S but generate a charge current which contributes to Joule heating, which in turn reduces the refrigerating effect.
To reduce sub-gap transport, an insulating barrier (I) is placed between the normal metal and the superconductor~\cite{Nahum1994,Bardas1995,Leivo1996}.
Such a barrier, however strongly limits the transfer of excitations above the gap, thus reducing the cooling.
It is therefore desirable to have a means of reducing the probability of Andreev reflection without using an insulating barrier. This may be achieved, for example, using magnetic effects
as in Refs.~\cite{Giazotto2002,Giazotto2006b,Ozaeta2012,Kawabata2013,Rouco2018}.

\begin{figure}[!tbh]
\centering
\includegraphics[width=0.9\columnwidth]{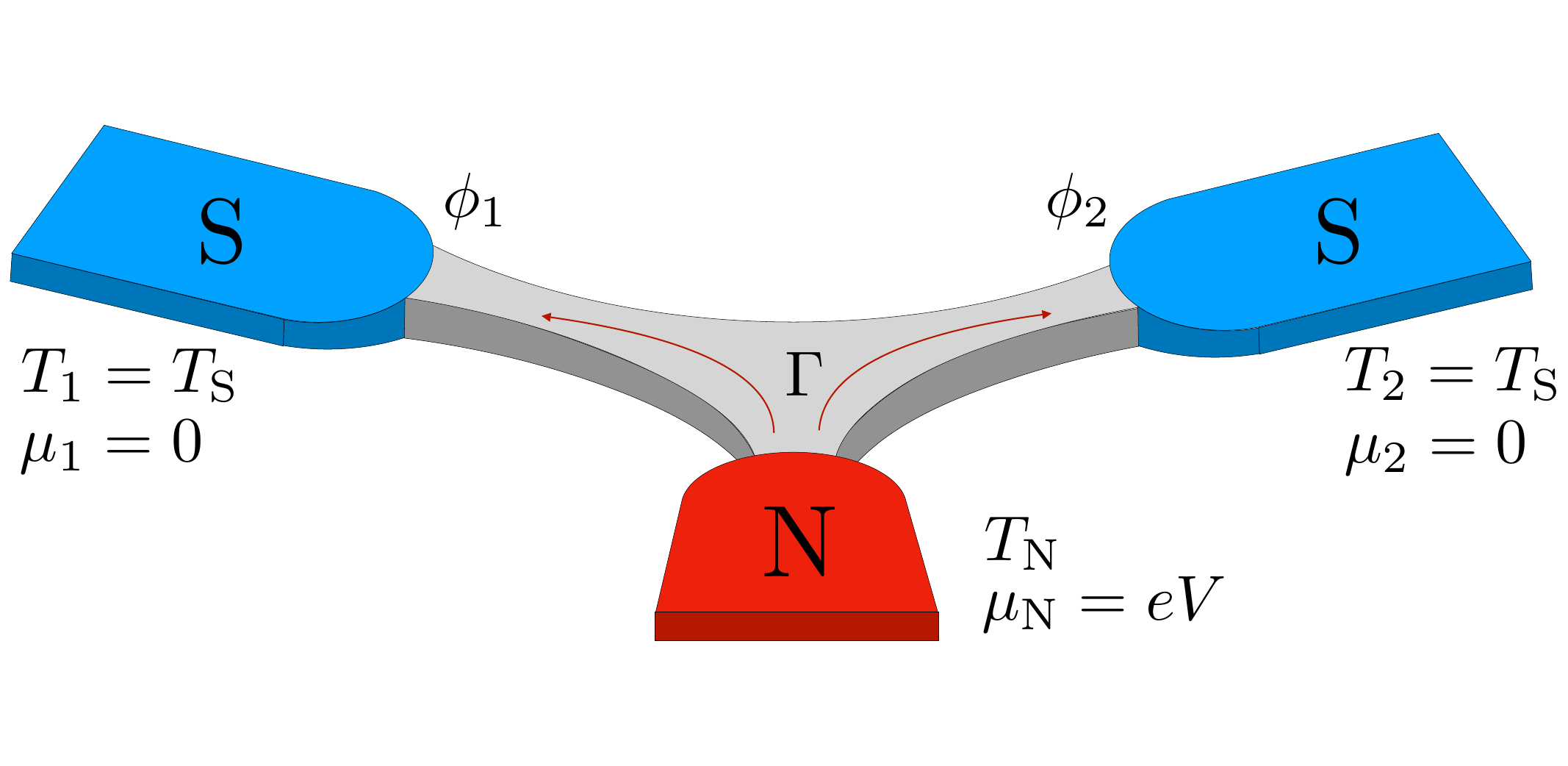}
\caption{Sketch of the Andreev interferometer. The normal (N) terminal, depicted in red, characterized by a chemical potential $\mu_{\rm N}$ and temperature $T_{\rm N}$, is located on the bottom, while the superconducting (S) terminals, in blue, both having a chemical potential of $\mu_1 = \mu_2 = 0$ and temperatures $T_1 = T_2 = T_{\rm S}$, are positioned on the top right and top left.
A phase difference $\Delta \phi = \phi_2 - \phi_1$ is established between the two superconductors. A beam splitter, in grey, with a transmission probability $\Gamma$ couples the N terminal to the S terminals.}
\label{interferometer_scheme_nodot}
\end{figure}

In the present paper, we propose a superconducting cooler based on a ballistic Andreev interferometer (AI) which enables a maximal cooling power per channel that, for realistic parameters, is five orders of magnitude larger with respect to standard NIS coolers.
The AI consists of a normal metal terminal, to be cooled down, attached to two superconducting (S) regions through two wires or a beam splitter, as in Fig.~\ref{interferometer_scheme_nodot}.
We exploit the destructive interference, occurring when the superconducting phase difference between the two S terminals equals $\pi$, to achieve the complete suppression of the Andreev reflection.
Even in the absence of insulating or magnetic elements, the charge current below the gap gets strongly suppressed without reducing the transmission of excitations above the gap, thus maximizing the heat extraction from the N terminal.
With respect to standard NIS coolers~\cite{Nahum1994,Bardas1995,Leivo1996} we find that the maximum cooling power occurs at a slightly larger temperature (0.32 $\Delta_0/k_B$ vs 0.25 $\Delta_0/k_B$).
Most strikingly, we find that, for realistic parameters, the maximal cooling power per channel can be up to five orders of magnitude larger.
As a result, we find that the electronic temperature which can be reached nearly equals the lower bound
determined by the maximum heat current which can flow in a two-terminal system~\cite{Pendry1983,Whitney2013}.
In addition, we find an approximate expression of such a bound valid for low phonon bath temperatures.
It should be stressed that the critical conditions are: i) the mean free path of the wires must be larger than the length of the wires (ballistic transport), and ii) the coupling to the two S regions must be symmetric.
We address the last condition by calculating the optimal cooling power for different lengths of the two wires and for a beam splitter with asymmetric arms. In the former case we also calculate  the minimal temperature achievable.
Finally, we suggest a few possible implementations of ballistic AI based on nanostructures realized with semiconductors, graphene and topological insulators.

The paper is organized as follows: In Section~\ref{Model}, we detail the model by defining the scattering matrix of the system, the heat current flowing out of the normal metal, and the energy balance equation needed to calculate the electronic temperature. In Section~\ref{Results}, we numerically calculate the cooling power for different sets of parameters and determine the cooling power at the optimal bias voltage. In Section~\ref{refr}, we calculate the electronic temperature and compare it with the theoretical minimum achievable, providing an approximate analytical expression.
In Section~\ref{dis}, we discuss the results by calculating the cooling power per channel, addressing the role of asymmetry in the arms of the beam splitter, and proposing possible implementations of an Andreev interferometer-based cooler. We also include several appendices where we detail some analytical calculations, such as the Andreev reflection amplitude in the case of arms of different lengths (Appendix~\ref{App-diffL}), the derivation of the theoretical minimum achievable temperature (Appendix~\ref{App-7}) and the matrix of Andreev reflection amplitudes in the multichannel case (Appendix~\ref{App-multi}). Conclusions are drawn in Section~\ref{conc}.

\section{Model}
\label{Model}
We consider an AI, schematically depicted in Fig.~\ref{interferometer_scheme_nodot}, which consists of a normal metal reservoir (N, red element in the center) connected to two superconducting terminals (S, blue elements on the sides) through a three-leg beam splitter (colored in grey).
The N terminal is characterized by a temperature $T_\mathrm{N}$ and chemical potential $\mu_\mathrm{N}=eV$, while the S terminals are kept at the same temperature $T_\mathrm{S}$ and grounded $\mu_1=\mu_2=0$. A superconducting phase difference $\Delta \phi = \phi_2 - \phi_1$ is set between them. 

The AI is described by its scattering matrix which can be calculated by composing~\cite{Datta1997} the scattering matrices of the beam splitter and of the two NS interfaces.
The beam splitter is assumed to be symmetric and described by the scattering matrix~\cite{Buttiker1984}
\begin{align}
\label{smat}
S_{\rm BS}=
\begin{pmatrix}
-s_1 \sqrt{1-2\Gamma} & \sqrt{\Gamma}  & \sqrt{\Gamma} \\
\sqrt{\Gamma}  & a & b \\
\sqrt{\Gamma}  & b & a
\end{pmatrix} ,
\end{align}
where $a=1/2(s_2+s_1\sqrt{1-2\Gamma})$, $b=1/2(-s_2+s_1\sqrt{1-2\Gamma})$, $0\leq\Gamma\leq 1/2$, $s_{1,2}=\pm$.
In Eq.~(\ref{smat}), $\Gamma$ is the probability of transmission from lead N into lead 1 and into lead 2 of the beam splitter. The time-reversed processes have the same probability. Similarly, $b$ is the probability amplitude of transmission from lead 1 to lead 2 and vice versa. On the other hand, $a$ is the probability amplitude of reflection at lead 1 and is equal to the amplitude for the reflection at lead 2.
Each ideal NS interface is described by the following perfect Andreev reflection amplitude
\begin{align}
\label{rhe}
r^{\rm he}_{\rm A} (E)= 
\exp (-i {\rm Arccos} ( E/|\Delta |) -i\phi),
\end{align}
for an electron to be reflected as a hole, and
\begin{align}
\label{reh}
r^{\rm eh}_{\rm A} (E)= \exp (-i {\rm Arccos} (E/|\Delta|) +i\phi),
\end{align}
for a hole to be reflected as an electron, where $\Delta$ is the superconducting order parameter and $\phi$ is the superconducting phase.
The total scattering probabilities that we are interested in are the Andreev reflection probability $R^{+-}(E)$ for a hole at energy $E$ to be reflected as an electron into lead N, the normal reflection probability $R^{++}(E)$ for an electron to be scattered back into lead N, and the total probability $T(E)$ for an electron injected from lead N to be transmitted into the superconducting leads as an electron or a hole. Remarkably, when $\Delta\phi=\pi$ the Andreev reflection probability $R^{+-}(E)$ vanishes for all energies as a result of destructive interference, independently of the value of $\Gamma$ (see Appendix~\ref{App-diffL}). Above the gap, transmission remains finite and close to 1 as long as $\Gamma$ is close to $1/2$, resulting in the maximization of energy transfer.

In this paper we are interested in the cooling power $J$, i.e.~the heat current flowing out of the N reservoir.
Within the scattering  approach, $J$ can be expressed as
\begin{equation}
\begin{split}
    J(V,T_{\rm N},T_{\rm S}) = \frac{2}{h} \int^\infty_{-\infty} dE \, (E - eV)  \{[1-R^{++}(E)]\\
    \times [f_{\rm N}^+(E) - f_{\rm S}(E)] -
     R^{+-}(E)[f_{\rm N}^-(E) - f_{\rm S}(E)]\},
\end{split}
\label{coolcurr}
\end{equation}
where $f_\mathrm{N}^\alpha(E)=\{\exp[(E-\alpha eV)/k_BT_{\rm N}]+1\}^{-1}$ is the Fermi function for electrons (for $\alpha=+1$) or holes (for $\alpha=-1$) in the normal reservoir, while $f_\mathrm{S}(E)=\{\exp(E/k_BT_{\rm S})+1\}^{-1}$ is the one for quasiparticles in the superconductor.
The factor $2$ in front accounts for the spin degeneracy.
We remark that the reflection probabilities also depend on the temperature of the superconducting terminal, through the temperature dependence of the superconducting gap $\Delta=\Delta (T_{\rm S})$. All the results presented in this work are calculated assuming the temperature dependence given by the BCS theory~\cite{Tinkham2004}.
We can now calculate the steady state temperature of the N reservoir by imposing the following energy balance equation~\cite{Roukes1985,Wellstood1994}
\begin{equation}
    J(\mu,T_\mathrm{N},T_\mathrm{ph}) + \Sigma \mathcal{V} (T_\mathrm{N}^5 - T_\mathrm{ph}^5) = 0,
    \label{baleq}
\end{equation}
where the second term is the rate of energy exchange between the electrons and the phonons. $\Sigma$ is the material-dependent electron-phonon coupling strength and $\mathcal{V}$ is the volume of the N electrode.
We have assumed that the temperature of the superconductor coincides with the one of the phononic bath, i.e~$T_\mathrm{S}=T_\mathrm{ph}$. 
Equation~(\ref{baleq}) can be numerically solved for $T_\mathrm{N}$ once the values of $V$ and $T_\mathrm{ph}$ are given.

\section{Cooling power}
\label{Results}

\begin{figure}[h]
\includegraphics[width=0.45\textwidth]{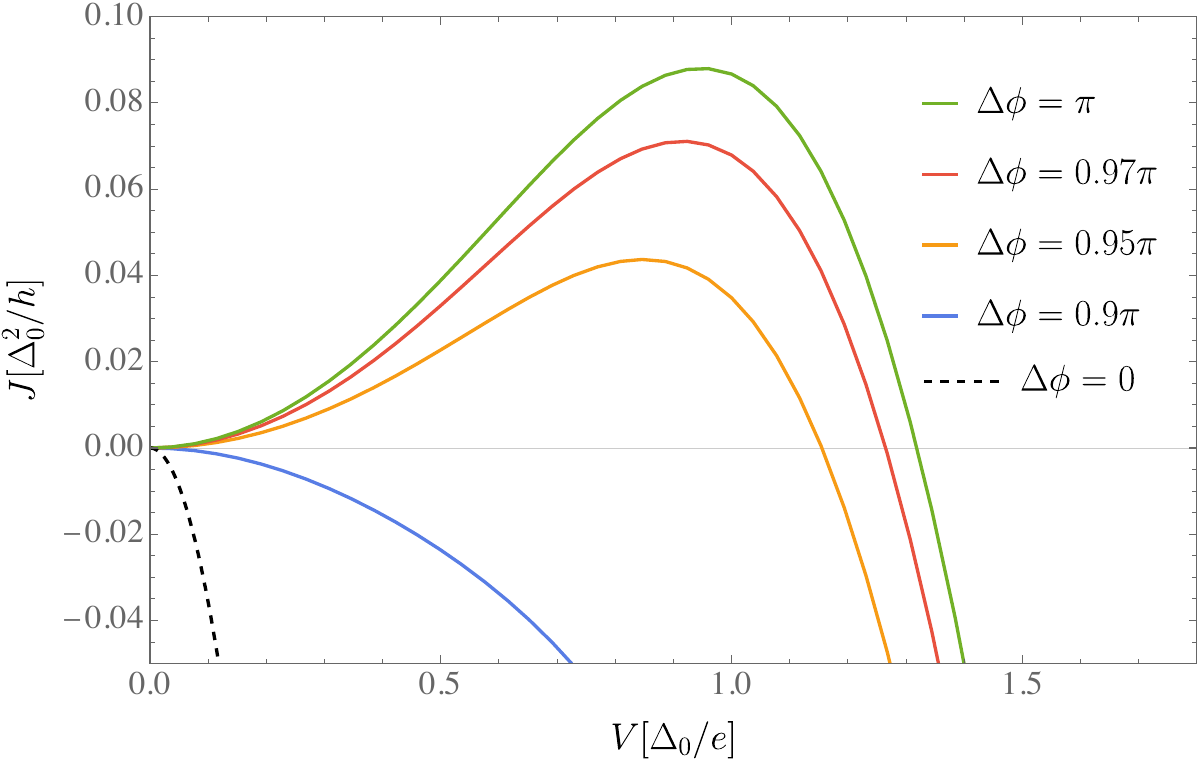}
\caption{Cooling power $J$ (measured in units of $\Delta_0^2/h$) plotted as a function of the voltage bias $V$ (in units of $\Delta_0/e$) for different values of the phase difference $\Delta \phi$ (see legend). We have used the following parameters: $\Gamma=0.5$, $s_1=+1$, $s_2=-1$ and $T_{\rm S}=T_{\rm N}=0.55 T_c$. The dashed black curve corresponds to the case where $\Delta \phi = 0$.}
\label{JvsV_nodot_T0.3}
\end{figure}

In Fig.~\ref{JvsV_nodot_T0.3} we plot the cooling power at equilibrium temperature, i.e. $J(V, T, T )$, as a function of the voltage, and for different values of $\Delta\phi$, calculated for $\Gamma=1/2$, $s_1=+1$, $s_2=-1$ and $T\equiv T_{\rm S}=T_{\rm N}=0.55 T_c$.
In particular, we consider the case of a transparent junction, which corresponds to the maximal coupling between N and S reservoirs ($\Gamma=1/2$). Indeed, the normal state resistance of the junction, given by $R_{\rm t}=h/(4e^2\Gamma)$, in this case, is the minimal attainable.
Remarkably, we find that the cooling power shows a positive peak at $V$ close to $\Delta_0/e$ (i.e.~heat current flows from the N reservoir to the S leads), at least for values of $\Delta\phi$ close to $\pi$, despite the fact that the junction is transparent.
Moreover, the shape of the curves in Fig.~\ref{JvsV_nodot_T0.3} is very similar to the ones relative to the ordinary NIS cooler~\cite{Giazotto2006}.
Note, however, that $J$ decreases very rapidly when $\Delta\phi$ departs from $\pi$ due to the cranking up of the Andreev reflection (the positive peak disappears for $\Delta\phi\simeq 0.92\pi$).
It is worth stressing that the case with $\Delta\phi=0$ (dashed black curve) is equivalent to the case of a transparent (single-channel) NS junction~\cite{Giazotto2006}, which presents a very large, negative, cooling power.
Finally, with $\Gamma$ decreasing, as expected, $J$ gets suppressed monotonically, and for all voltages, but still remains positive in roughly the same voltage range for $\Delta\phi=\pi$ (not shown).

\begin{figure}[t]
    \centering
    \begin{overpic}
        [width=0.82\columnwidth]{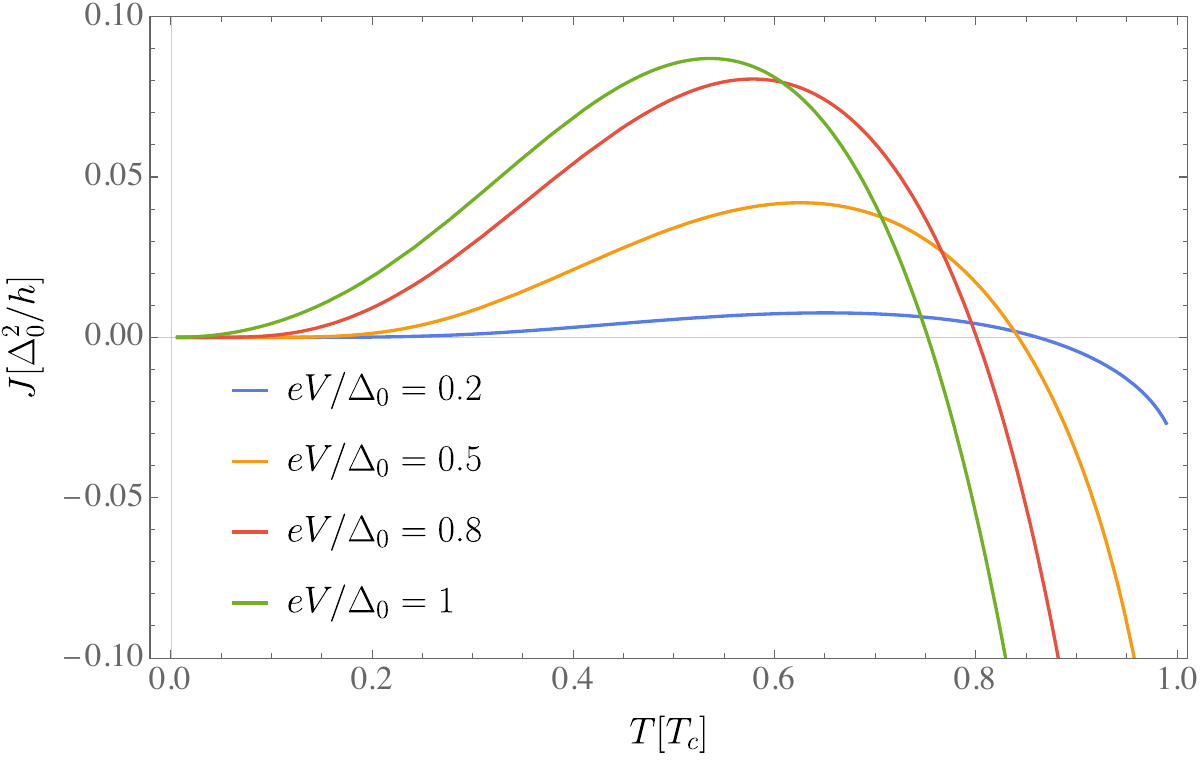}\put(-10, 115){$(a)$}
        \put(30,95){\linethickness{0.02mm}\color{black}\polygon(0,0)(38,0)(38,25)(0,25)}%
        \put(33,110){$\Delta \phi = \pi$}%
        \put(34,100){$\Gamma = 0.5$}%
    \end{overpic}
    \hfill
    \begin{overpic}
        [width=0.86\columnwidth]{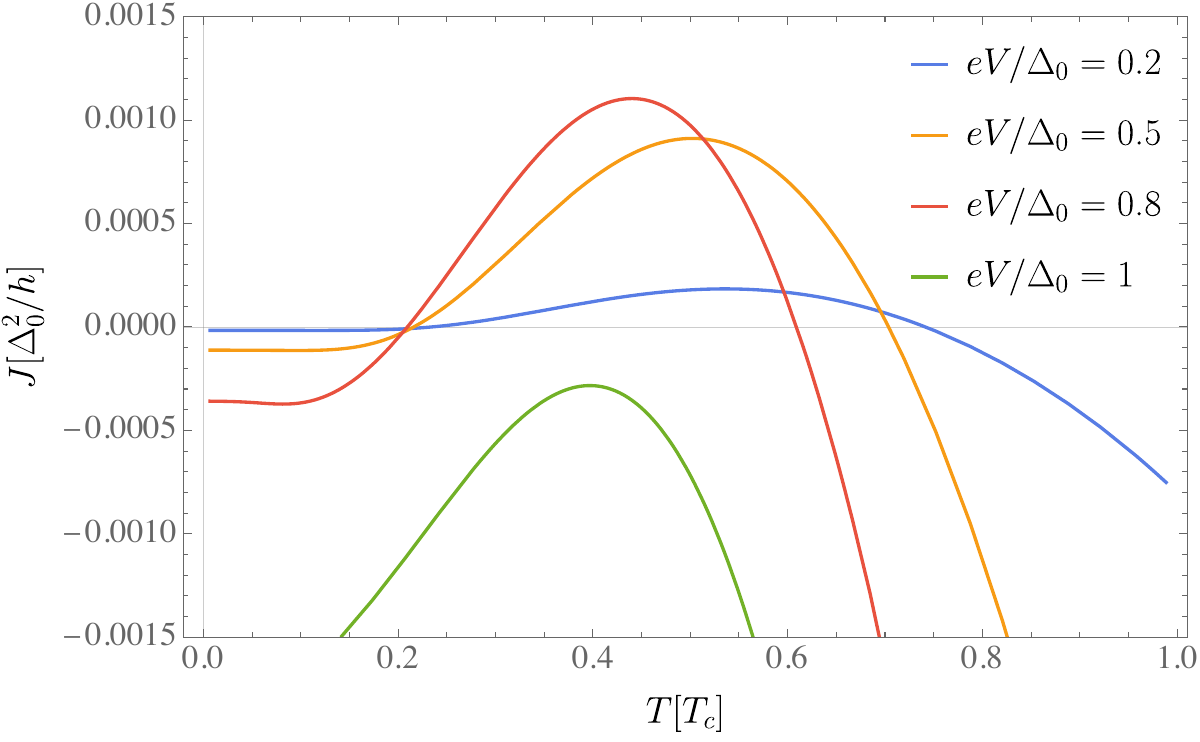}\put(-3, 120){$(b)$}
        \put(37,95){\linethickness{0.02mm}\color{black}\polygon(0,0)(42,0)(42,25)(0,25)}%
        \put(40,110){$\Delta \phi = 0$}%
        \put(40,100){$\Gamma = 10^{-2}$}%
    \end{overpic}
    \hfill
    \begin{overpic}
        [width=0.9\columnwidth]{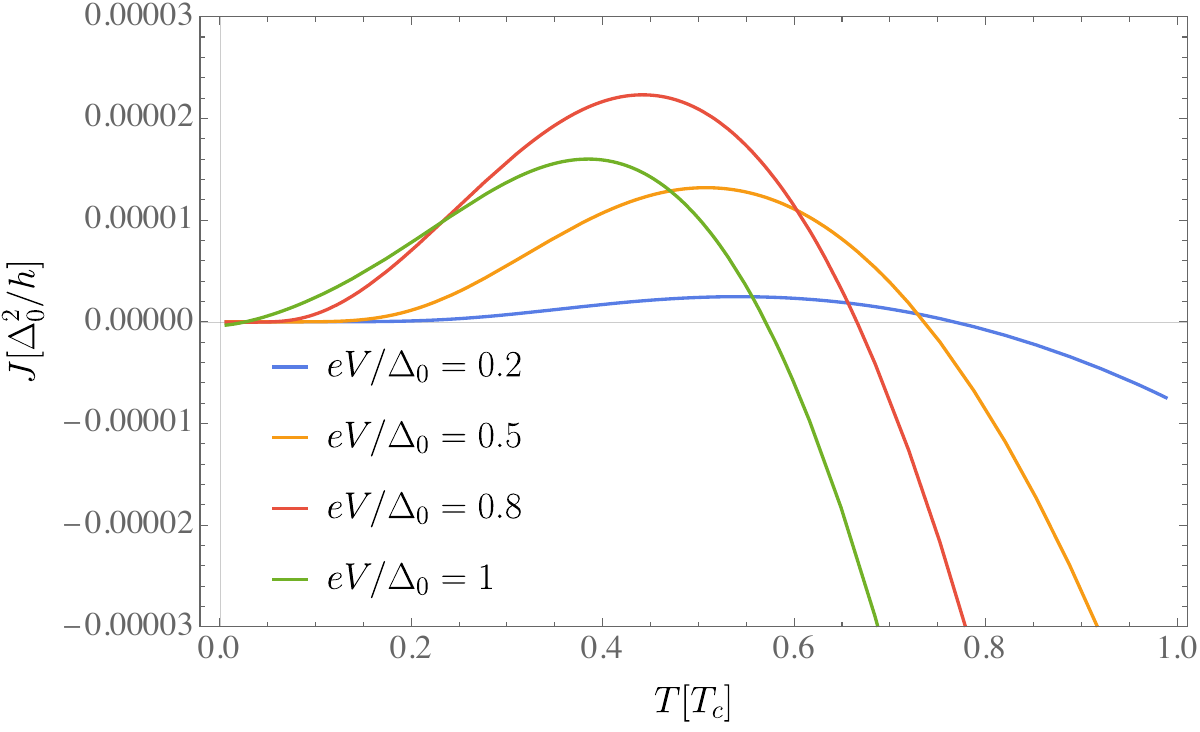}\put(1, 120){$(c)$}
        \put(42,100){\linethickness{0.02mm}\color{black}\polygon(0,0)(42,0)(42,25)(0,25)}%
        \put(45,115){$\Delta \phi = 0$}%
        \put(45,105){$\Gamma = 10^{-4}$}%
    \end{overpic}
    \caption{Cooling power $J$  (measured in units of $\Delta_0^2/h$) plotted as a function of the equilibrium temperature (measured in units of $T_c$) for several values of the bias voltage $V$. Panel (a) refers to the case with $\Gamma=0.5$ and $\Delta \phi = \pi$, i.e. $R^{+-}=0$ for every value of the energy. Panel (b) refers to the case with $\Delta \phi = 0$ and $\Gamma=0.01$.
    Panel (c) refers to the tunneling limit case with $\Delta \phi = 0$ and $\Gamma=10^{-4}$, which is equivalent to a single-channel NIS cooler.
We have used the following parameters: $s_1=+1$, $s_2=-1$ and $T_{\rm S}=T_{\rm N}=0.55 T_c$.}
    \label{JvsT}
\end{figure}

\begin{figure}[h]
\includegraphics[width=0.45\textwidth]{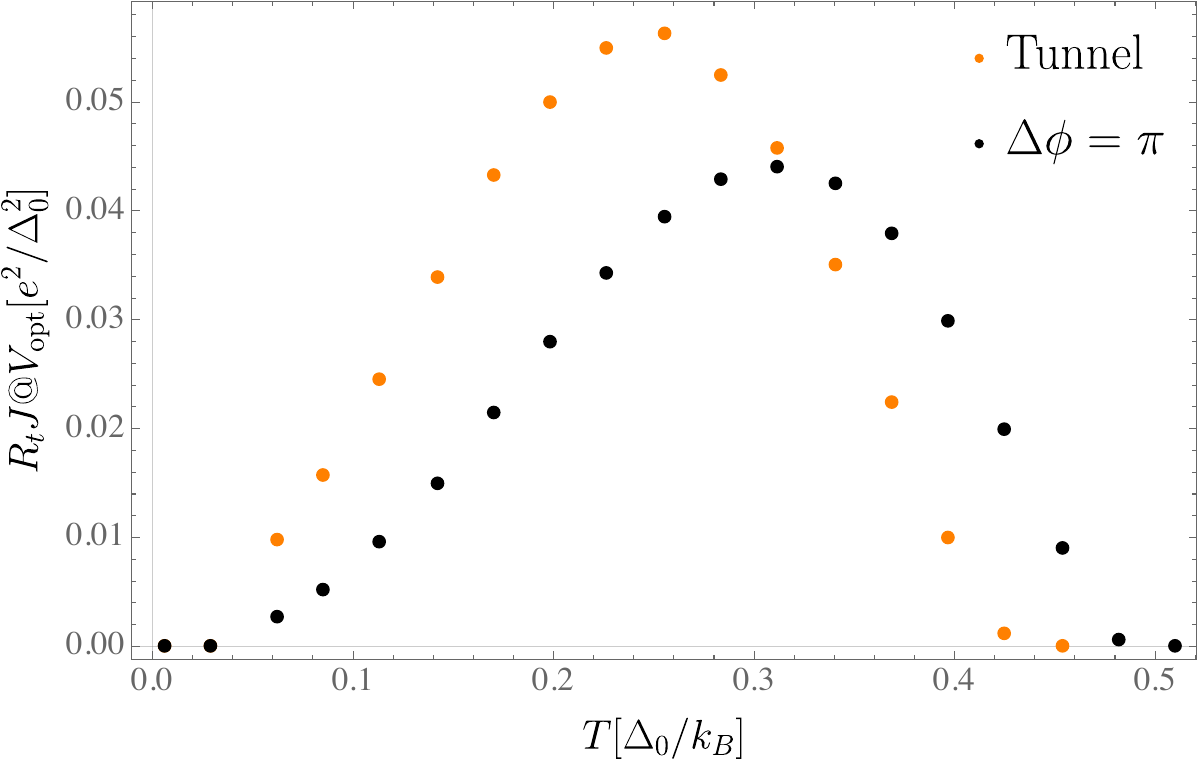}
\caption{Normalized cooling power $R_tJ$, measured in units of $e^2/\Delta_0^2$, calculated at the optimal bias $V_{\rm opt}$ as a function of the equilibrium temperature $T$. $R_t$ is the resistance of the junction in the normal state; see text. Black dots correspond to $\Delta \phi = \pi$ and $\Gamma=0.5$. For comparison, we also report orange dots corresponding to the NIS cooler ($\Delta \phi = 0$ and $\Gamma=10^{-4}$).
}
\label{JoptvsT_PivsTunnel_Gn}
\end{figure}

It is now insightful to consider the behavior of the cooling power at equilibrium temperature $J(V,T,T)$ as a function of $T$. This is illustrated in Fig.~\ref{JvsT} for $\Delta\phi=\pi$ [panel (a)] and $\Delta\phi=0$ [panel (b) and (c)], and for different values of voltage $V$ and transmission probability $\Gamma$, specifically $\Gamma=1/2$ (transparent case) in panel (a), $\Gamma=0.01$ in panel (b), and $\Gamma=10^{-4}$ (tunneling case) in panel (c).
Our aim here is to compare our proposal with the performance of the well known paradigmatic (single-channel) NIS cooler.
For a transparent junction, Fig.~\ref{JvsT}(a) shows that the cooling power remains positive in a wide range of temperatures, from 0 to $\overline{T}$, with $\overline{T}$ decreasing with increasing $V$. Moreover, for all values of $V$ considered, $J(V,T,T)$ exhibits a positive peak, whose height and position depend on the voltage. The largest peak occurs for $V\simeq \Delta_0/e$.
This behavior can be contrasted to
the single-channel NIS cooler scenario, which we implement by choosing a small value of $\Gamma$ and setting $\Delta\phi=0$.
Indeed, in this case, the cooling power $J(V,T,T)$, as shown in Fig.~\ref{JvsT}(b) for $\Gamma = 0.01$, still exhibits positive peaks (although they are of small absolute height), but it becomes negative for temperatures below approximately 0.2 $T_c$.
This is a consequence of Joule heating occurring because of the presence of a finite (though small) Andreev reflection below the gap.
Note that the curve relative to $eV=\Delta_0$ is negative for all values of $T$.
If now we further reduce $\Gamma$ and set $\Gamma = 10^{-4}$, see Fig.~\ref{JvsT}(c), we eventually reach the tunneling limit of the single-channel NIS cooler setup~\footnote{One can check that the curves in Fig.~\ref{JvsT}(c), once normalized to the normal state resistance of the junction $R_{\rm t}$, exactly match the ones where the cooling power is calculated with the tunneling formula, see Ref.~\cite{Giazotto2006}.}.
As expected, the maximum cooling power further decreases, by many orders of magnitude for this choice of parameters, since we are decreasing the transmission probability of the NIS junction.
As we will see below, this reflects in an extremely small cooling power per open channel. The shape of these curves somehow resembles the ones plotted in panel (a), apart from a shift of all the curves towards lower temperatures, and the largest peak now occurring at $V\simeq 0.8\Delta_0/e$.
For a given temperature $T$, we define the
optimal bias $V_{\rm opt}$ as the value of $V$ which maximizes the cooling power.
In Fig.~\ref{JoptvsT_PivsTunnel_Gn} we plot the cooling power at the optimal bias $J(V_{\rm opt},T,T)$, normalized to the normal state resistance $R_t=h/(4e^2 \Gamma)$, as a function of $T$.
We consider two cases: $\Delta \phi = \pi$ and $\Gamma = 1/2$ (black dots), and $\Delta \phi =0$ and $\Gamma = 10^{-4}$ (orange dots), i.e.~the
NIS cooler in the tunneling limit~\cite{Giazotto2006}. 
The two curves are fairly similar, the  NIS cooler though having a higher peak (0.06 vs 0.04 $\Delta_0^2G_{\rm N}/e^2$) occurring at smaller temperatures (0.25 vs 0.3 $\Delta_0/k_B$).

\section{Refrigeration}
\label{refr}
\begin{figure}[h]
\includegraphics[width=0.45\textwidth]{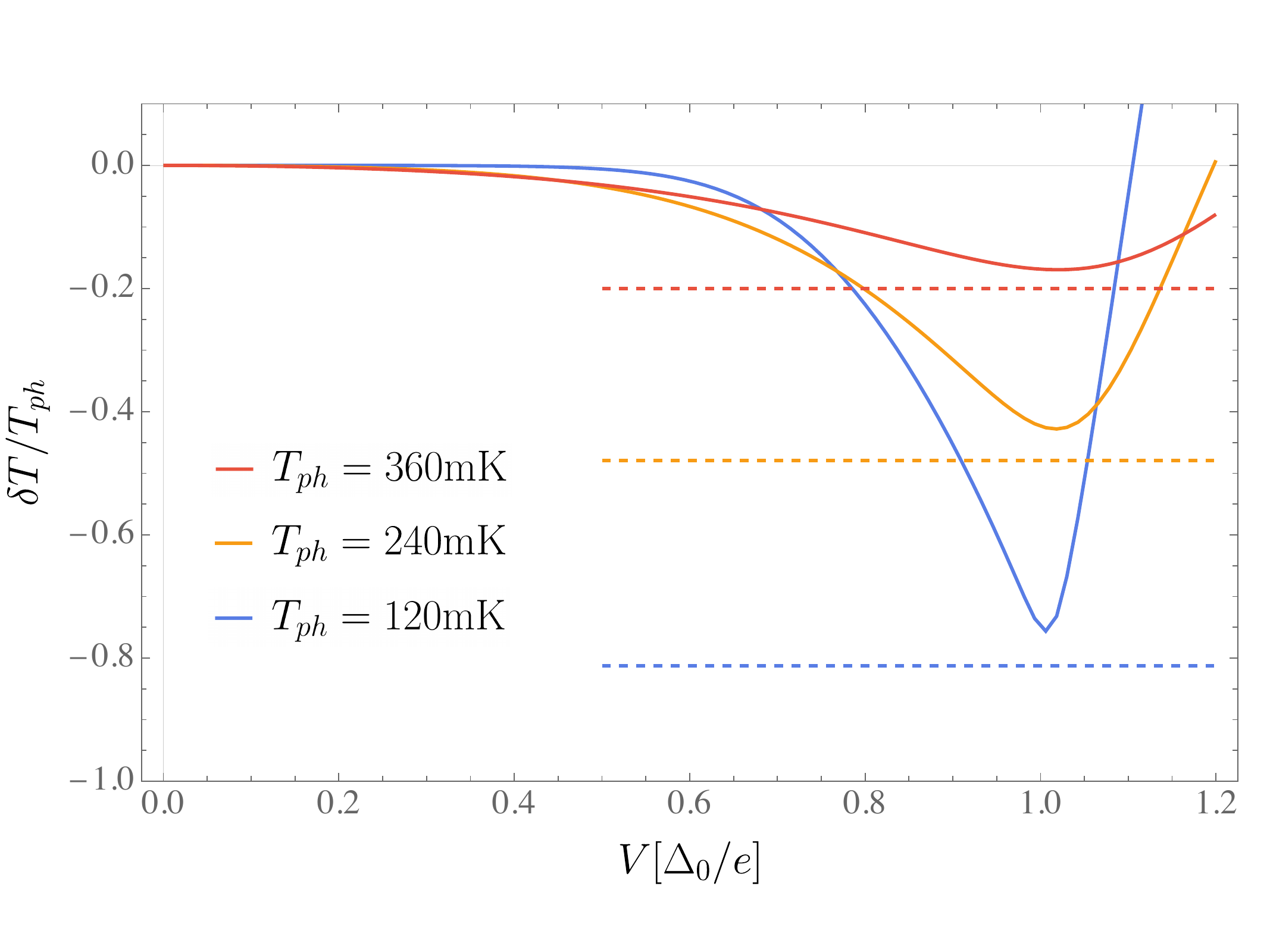}
\caption{Refrigeration. Relative deviation $\delta T/T_{\rm ph}=(T_\mathrm{N}-T_\mathrm{ph})/T_{\rm ph}$ of the equilibrium temperature $T_\mathrm{N}$ from the phonon temperature $T_\mathrm{ph}$ plotted as a function of the voltage bias $V$ (in units of $\Delta_0/e$). The various curves correspond to different values of $T_{\rm ph}$. The dashed curves refer to the minimum achievable temperature $T_\mathrm{N,min}$ (normalized to $T_{\rm ph}$)} calculated using the bound in Eq.~\eqref{whitney_bound}.
The parameters used are: $\Delta \phi = \pi$, $\Gamma=0.5$, $\Sigma= 10^9$ WK$^{-5}$m$^{-3}$, $\mathcal{V} \simeq 10^{-20}$ m$^3$ and critical temperature $T_c = 1.2$ K (for aluminum).
\label{deltaTvsV_nodot_Al}
\end{figure}

Now, to assess the actual ability of the AI with a transparent junction ($\Gamma=1/2$) to refrigerate the electrons in the metallic reservoir N, we solve the energy balance equation~(\ref{baleq}) setting the phase difference $\Delta\phi=\pi$ and also accounting for the temperature dependence of the superconducting gap $\Delta(T_{\rm S})$, as already mentioned.
We set $\Sigma= 10^9$ W $\mathrm{K}^{-5}\mathrm{m}^{-3}$ as a typical value for a metal~\cite{Giazotto2006,Giazotto2007}, $\mathcal{V} \simeq 10^{-20}$ $\mathrm{m}^3$ and $T_c=1.2 \ \rm K$ for aluminum.
In Fig.~\ref{deltaTvsV_nodot_Al} we plot the deviation of the equilibrium temperature $T_\mathrm{N}$ from the phonon temperature $T_\mathrm{ph}$, i.e.~$\delta T=T_\mathrm{N}-T_\mathrm{ph}$ normalized to  $T_\mathrm{ph}$, as a function of the voltage $V$ for different values of the phonon temperature, i.e.~$T_\mathrm{ph}=120$ mK (blue curve), $T_\mathrm{ph}=240$ mK (yellow curve) and $T_\mathrm{ph}=360$ mK (red curve).
By increasing the voltage, $\delta T$ decreases below zero and reaches a minimum for $V\sim \Delta_0/e$.
Analogously to the case of the well-known NIS cooler, the refrigeration effect is more effective for smaller temperatures.
Indeed, for $T_\mathrm{ph}=120$ mK the reduction is about 75\%, which corresponds to a
minimum value of $T_{\rm N}=30$ mK, while for $T_\mathrm{ph}=240$ mK the 
reduction is about 43\%, which corresponds to a
minimum value of $T_{\rm N}=138$ mK,
and for $T_\mathrm{ph}=360$ mK 
it is about 17\%, which corresponds to a
minimum value of $T_{\rm N}=300$ mK.
Note that the maximum value of $\delta T$ obtained is of the same order as the one measured using conventional NIS refrigerators, see for example~Refs.~\cite{Giazotto2006,Leivo1996}.
However, at odds with the NIS cooler, by increasing the phonon temperature the position of the minimum slightly moves to lower values of $V$ and the dip in the curve widens.

We now show that our AI cooler allows us to nearly reach the maximum refrigeration.
This is represented by the minimal possible temperature which is set by the maximum cooling power given by the upper bound on the heat current flowing out of the colder reservoir in a two-terminal system, namely~\cite{Pendry1983,Whitney2013,whitney2014most,whitney2015finding}
\begin{equation}
    J \leq J_\mathrm{bound} \equiv \frac{N \pi^2 (k_\mathrm{B}T_\mathrm{N})^2}{12h}.
    \label{whitney_bound}
\end{equation} 
Here $N$ is the number of transverse channels in the colder (normal) lead which, in our case, is equal to 2 due to the spin degeneracy.
Since this bound does not depend on either the temperature of the hotter reservoir or on the voltage bias, we can substitute $J_\mathrm{bound}$ into the energy balance equation~(\ref{baleq}) to obtain the minimum achievable temperature of the normal terminal $T_\mathrm{N,min}$ for a given phonon temperature $T_\mathrm{ph}$.
In Fig.~\ref{deltaTvsV_nodot_Al}, the values of $T_\mathrm{N,min}$, normalized to  $T_\mathrm{ph}$, are represented by dashed horizontal lines with colors corresponding to the different phonon temperatures.
For example, at $T_\mathrm{ph} = 120$ mK we find $T_\mathrm{N,min} \simeq 22$ mK, while our cooler reaches $T_\mathrm{N} \simeq 30$ mK.
At $T_\mathrm{ph} = 240$ mK we find $T_\mathrm{N,min} \simeq 125$ mK, while our cooler reaches $T_\mathrm{N} \simeq 138$ mK. At $T_\mathrm{ph} = 360$ mK we find $T_\mathrm{N,min} \simeq 288$ mK, while our cooler reaches $T_\mathrm{N} \simeq 300$ mK.
Remarkably, in all cases, the minimum temperature achieved $T_\mathrm{N}$ is close to the theoretical minimum $T_\mathrm{N,min}$.
Furthermore, for the normalized maximum cooling power one finds $R_t J_{\rm bound}\simeq$ 0.07 $\Delta_0^2/e^2$ by choosing $\Gamma=1/2$ and $T_N=0.3 \Delta_0/k_B$ (the conditions which give the maximum in Fig.~\ref{JoptvsT_PivsTunnel_Gn}).
It is interesting to notice that this value is only slightly above the maximum value of $R_t J$ which one can read from Fig.~\ref{JoptvsT_PivsTunnel_Gn}.

For low $T_\mathrm{ph}$ (roughly smaller than $0.4 T_\mathrm{C}$), we find the following approximation for the minimum achievable temperature of the normal terminal 
\begin{equation}
        \frac{T_\mathrm{N,min}}{T_\mathrm{C}} \simeq \sqrt{\frac{12 h \Sigma}{\pi^2 k_\mathrm{B}^2} \frac{ \mathcal{V} T_\mathrm{C}^3}{N} \left(\frac{T_\mathrm{ph}}{T_\mathrm{C}}\right)^5},
        \label{whitney_approx}
\end{equation}
see Appendix~\ref{App-7} for details.
We notice that $T_\mathrm{N,min}$ decreases as $1/\sqrt{N}$ with the number of channels $N$ and increases as $\sqrt{\mathcal{V}}$ with the volume of the conductor.
Remarkably, this means that a small increase in the number of channels would not lead to a significant improvement in the cooling.
Note that for our choice of parameters we find that the quantity $\frac{12 h \Sigma}{\pi^2 k_\mathrm{B}^2} \frac{\mathcal{V}}{N} T_\mathrm{C}^3\simeq 35$, so that $T_\mathrm{N,min}/T_c\simeq 6 (T_\mathrm{ph}/T_c)^{5/2}$.

\section{Discussion}
\label{dis}

\subsection{Cooling power per channel}

Thus far, we have assumed that the AI supports a single channel. It is important to emphasize that, for $\Delta \phi = \pi$, destructive interference also occurs in the case of multichannel leads, as demonstrated in Appendix~\ref{App-multi}. This means that our cooling scheme remains effective in this scenario and can actually achieve greater cooling power.
Consequently, it becomes relevant to discuss the cooling power per channel.
As previously mentioned in the discussion of Fig.~\ref{JvsT}, the primary advantage of our AI cooler is its substantial cooling power per channel. This advantage can be quantified by comparing it to that of the NIS cooler. Specifically, we can use the fact that the maximum value of the normalized cooling power $R_tJ_{\rm max}$ (obtained by maximizing over temperature and voltage) is actually similar in both cases—the NIS and AI coolers.
As shown in Fig.~\ref{JoptvsT_PivsTunnel_Gn}, one has that $(R_tJ_{\rm max})_{\rm NIS}\simeq 1.3(R_tJ_{\rm max})_{\rm AI}$.
From this relation, by multiplying and dividing by the number of open channel in the two different cooling schemes, we arrive at
\begin{equation}
  \frac{J_{\rm max, AI}}{N_{\rm AI}} \simeq \left(\frac{R_{t,{\rm NIS}} N_{\rm NIS}}{1.3\; R_{t,{\rm AI}}N_{\rm AI}}\right) \frac{J_{\rm max, NIS}}{N_{\rm NIS}}.
\end{equation}
By substituting realistic values~\cite{Giazotto2006b} of number of channels and resistance we obtain
\begin{equation}
    \left ( \frac{J_{\rm max}}{N} \right )_{\rm AI} \simeq 3 \times 10^5 \left ( \frac{J_{\rm max}}{N} \right )_{\rm NIS},
\end{equation}
which means that the cooling power per channel for our AI cooler is five orders of magnitude greater than for the NIS cooler.

\subsection{Non-ideal beam splitter: arms of different length and asymmetric scattering matrix}
\label{The case of different arms}

As we previously discussed, the beam splitter must be symmetric to achieve perfect destructive interference, thereby minimizing Andreev reflection.
In view of an experimental implementation, we now examine the effects of an asymmetric beam splitter configuration on the maximal cooling power and on the refrigeration. Specifically, we analyze the consequences of introducing a length difference $\delta L=L_2-L_1$ between the two beam splitter arms.
To quantify this effect, we calculate the corresponding scattering coefficients entering Eq.~(\ref{coolcurr}) following the methodology in Ref.~\cite{Blasi2023}.
Our approach explicitly accounts for the distinct dynamical phases accumulated when traversing the two beam splitter arms (see the details of the calculation in Appendix~\ref{App-diffL}).

\begin{figure}[h!]
    \centering
    \begin{overpic}
        [width=0.42\textwidth]{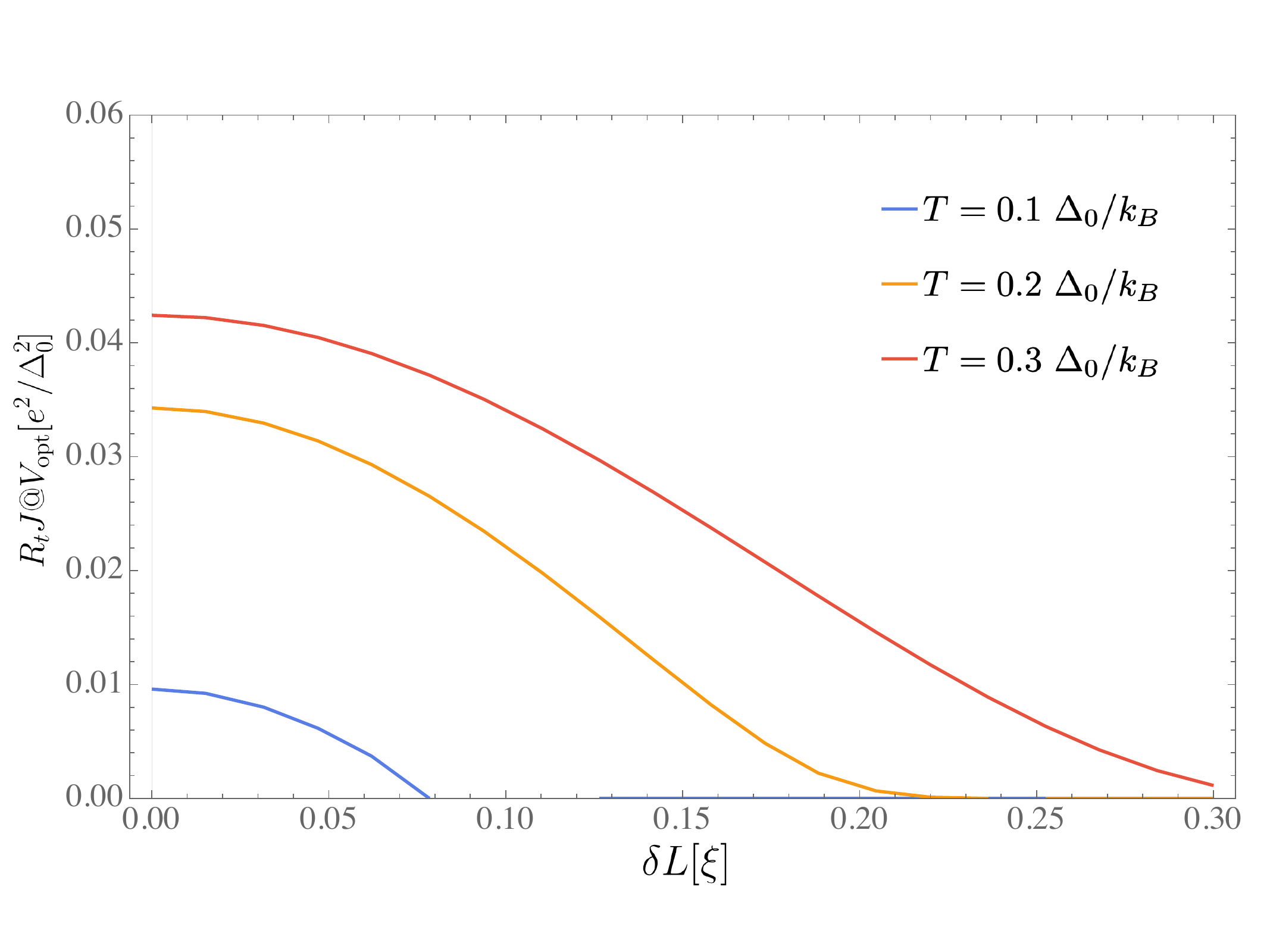}\put(-5, 150){$(a)$}
    \end{overpic}
    \hfill
    \begin{overpic}
        [width=0.42\textwidth]{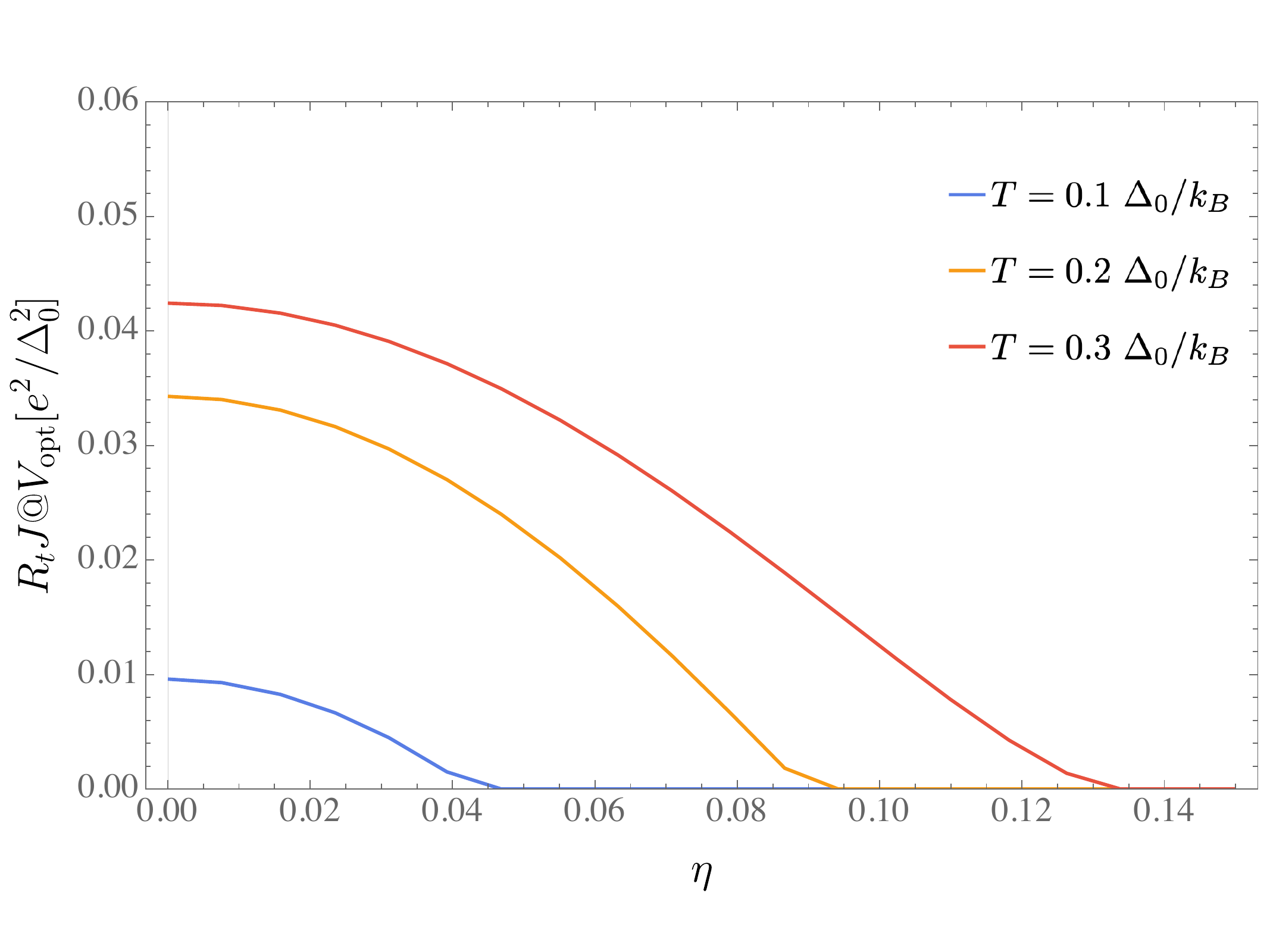}\put(-5, 150){$(b)$}
    \end{overpic}
    \hfill
    \begin{overpic}
        [width=0.42\textwidth]{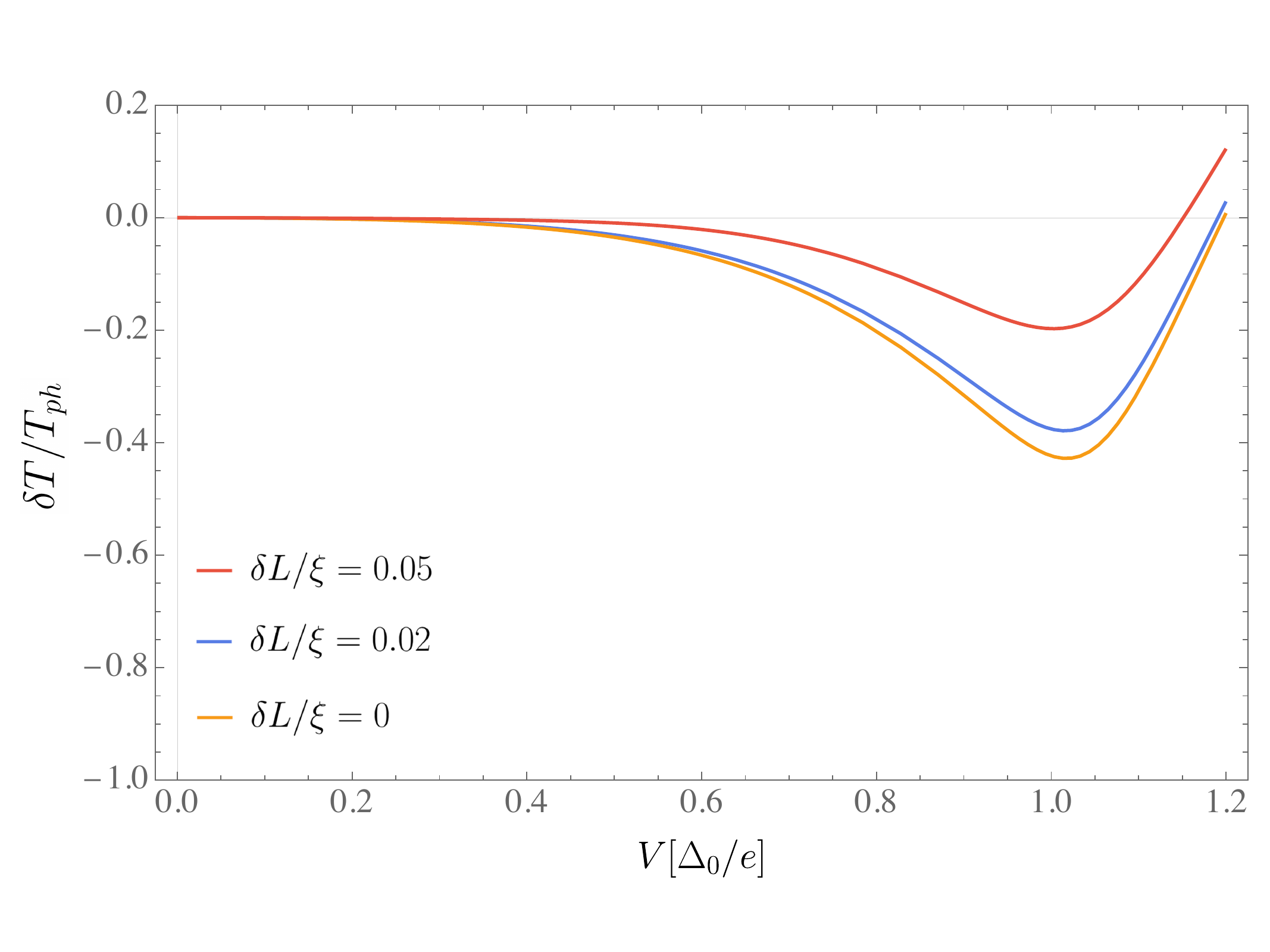}\put(-5, 150){$(c)$}
    \end{overpic}
    \caption{Non-ideal beam splitter: effects of different arm lengths
    $\delta L=L_2-L_1$ and of an asymmetric scattering matrix. Panel (a) and (b) refer to the normalized cooling power at the optimal bias $R_tJ(V_{\rm opt},T,T)$ as a function, respectively, of $\delta L$ and of $\eta$, for three different values of temperature $T$.
Panel (c) shows the relative temperature deviation $\delta T/T_{\rm ph}=(T_\mathrm{N}-T_\mathrm{ph})/T_{\rm ph}$} plotted as a function of the voltage bias $V$ (in units of $\Delta_0/e$) for three different values of $\delta L=L_2-L_1$ and for $T_{\rm ph} = 240$ mK. We set $\eta=0$, $\Delta \phi = \pi$, and $L_1 = 0.7 \xi$, where $\xi \simeq 1 \mu$m. Other parameters are as in Fig.~\ref{deltaTvsV_nodot_Al}.
    \label{deltaTNvsV_Tph0.2Al_vsdL}
\end{figure}

For $\Delta \phi = \pi$, in Fig.~\ref{deltaTNvsV_Tph0.2Al_vsdL}(a) we plot the normalized cooling power at the optimal bias $R_tJ(V_{\rm opt},T,T)$ as a function of the length difference $\delta L$, measured in units of the superconducting coherence length $\xi=\hbar v_F/ \Delta$, for three different values of temperature $T$ and for $L_1=0.7\xi$. The blue curve refers to $T=0.1 \Delta_0/k_B$, the yellow curve to $T=0.2 \Delta_0/k_B$, and the red curve to $T=0.3 \Delta_0/k_B$.
The three curves, as expected, are decreasing functions of $\delta L$, and start from different values of cooling power at $\delta L=0$ (see also Fig.~\ref{JoptvsT_PivsTunnel_Gn}).
In particular, while for a small temperature (blue curve) $R_tJ(V_{\rm opt},T,T)$ goes to zero already at $\delta L/\xi\simeq 6$\%, for $T=0.3\Delta_0/k_B$, $R_tJ(V_{\rm opt},T,T)$ remains finite up to a large value of length difference, namely $\delta L/\xi\simeq 30$\%.
Remarkably, for all three cases, $R_tJ(V_{\rm opt},T,T)$ depends only weakly on $\delta L$ up to $\delta L/\xi\simeq 3$\%, making the maximal cooling power fairly insensitive to small asymmetries.

To assess the dependence of the cooling on the asymmetry of the scattering matrix of the beam splitter itself, in Fig.~\ref{deltaTNvsV_Tph0.2Al_vsdL}(b) we plot, for $\Delta \phi = \pi$, the normalized cooling power at the optimal bias as a function of the degree of asymmetry $\eta$.
The latter is defined in terms of the transmission components of the scattering matrix of the beam splitter as
\begin{equation}
    \eta=\frac{\vert(S_{\rm BS})_{12}\vert^2-\vert(S_{\rm BS})_{13}\vert^2}{\vert(S_{\rm BS})_{12}\vert^2+\vert(S_{\rm BS})_{13}\vert^2} .
\end{equation}
The three curves in Fig.~\ref{deltaTNvsV_Tph0.2Al_vsdL}(b), corresponding to different values of temperatures as in panel (a), have very similar behaviors as compared with the ones in panel (a).
Also here, when the degree of asymmetry varies by up to a few percent, $R_tJ(V_{\rm opt},T,T)$ changes quite weakly.

We also calculate the temperature deviation $\delta T$ resulting from a length difference $\delta L$. Figure~\ref{deltaTNvsV_Tph0.2Al_vsdL}(c) shows, for $\Delta \phi = \pi$, $\delta T$ as a function of the applied voltage $V$ for a few values of $\delta L$.
The yellow curve represents the ideal scenario with $\delta L = 0$, and is referenced in Fig.~\ref{deltaTvsV_nodot_Al} with the same color. The blue curve, corresponding to a length difference of 2\% of the superconducting coherence length $\xi$, closely tracks the ideal curve. The temperature reduction is dramatically suppressed—effectively halved—only
when $\delta L$ increases to 5\% of $\xi$.
However, $\delta T$, as opposed to the maximum cooling power, appears to be sensitive to the geometric asymmetry of the beam splitter. This is due to the fact that the effect of a finite $\delta L$ on the cooling power is more effective at small temperatures and at large voltages, i.e.~in the conditions where maximal refrigeration occurs, see Fig.~\ref{deltaTvsV_nodot_Al}.
At higher temperatures we expect the refrigeration to be more stable against $\delta L$.

\subsection{Implementations}

\begin{figure}[h]
\includegraphics[width=0.4\textwidth]{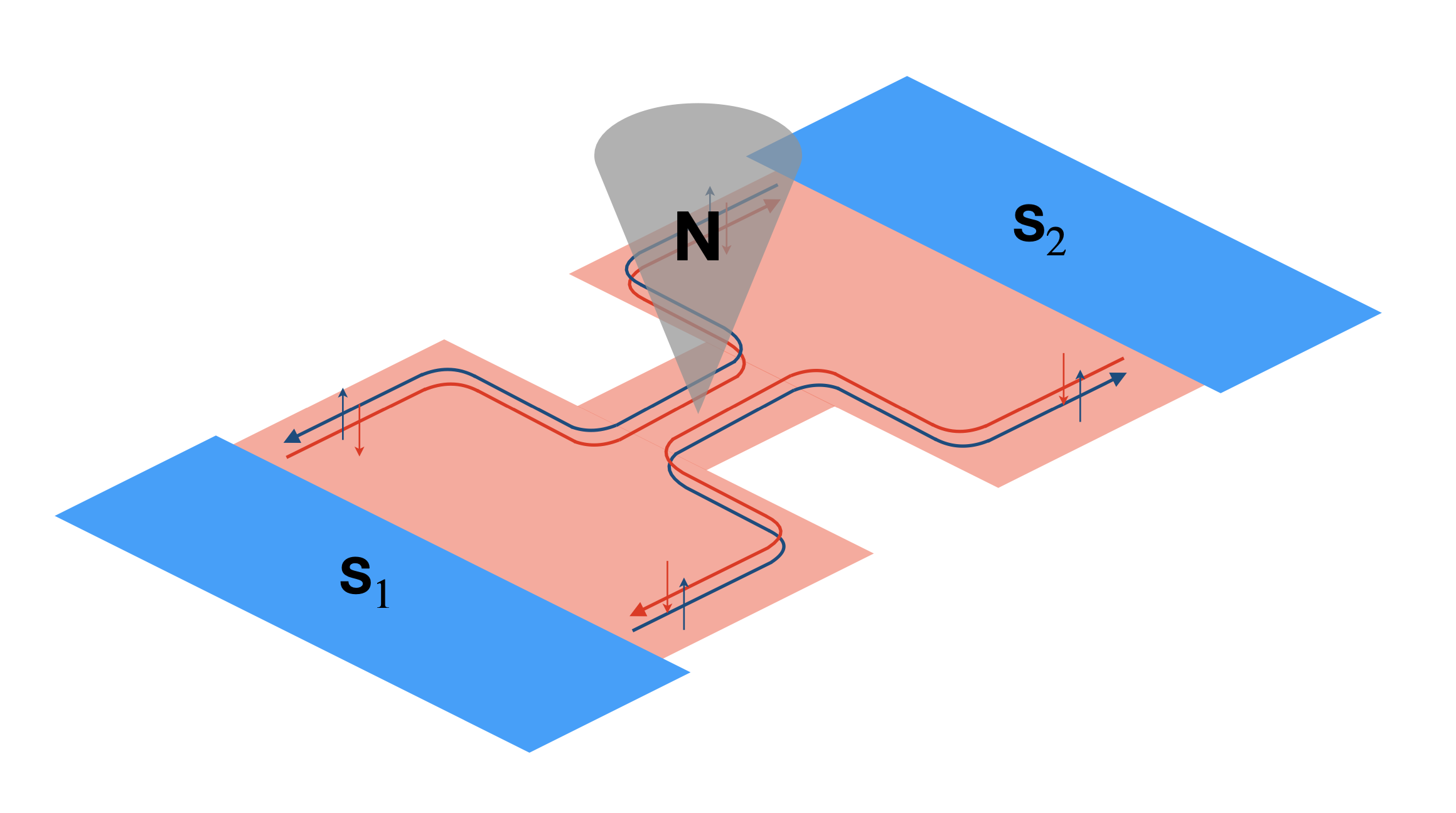}
\caption{Scheme of the implementation of a ballistic Andreev interferometer based on a 2D topological insulator (pink region). Blue and red lines represent helical edge states propagating in opposite directions: blue lines correspond to spin-up edge states, and red lines correspond to spin-down edge states. The N terminal is shown as a tip coupled to all the helical edge states, while the S terminals are colored blue.}
\label{sch-2DTI}
\end{figure}

We envisage various scenarios regarding the possible implementations of a ballistic Andreev interferometer~\cite{Claughton1996,Melin2024}. The first scenario is based on two-dimensional electron gases realized in semiconducting heterostructures. For example, using high-mobility GaAs/AlGaAs~\cite{Yamamoto2012}, analogous to what was proposed for an Aharonov-Bohm ring in Ref.~\onlinecite{Haack2019}, or using InAs~\cite{Amado2014} or InGaAs, similar to the proposal in Ref.~\onlinecite{Blasi2023}.
In those ballistic structures, Schottky barrier-free contacts with metals can typically be realized, maximizing transparency. These configurations allow for detailed control of the beam splitter through side and top gates. Furthermore, we expect that a limited number of channels will be involved in transport~\cite{Haack2019,Blasi2023}.
The second scenario is based on ballistic graphene/superconductor junctions~\cite{Cheng2009,Pandey2019,Takagaki2023}, which can be exploited to realize ultraclean graphene Andreev interferometers as in Ref.~\cite {Rashid2024}.
A third scenario is based on two-dimensional topological Josephson junctions~\cite{Sothmann2017} which incorporate an additional normal probe (see for example Refs.~\cite{Blasi2020,Blasi2020b}) and a constriction in the weak link, as sketched in Fig.~\ref{sch-2DTI}.
The normal probe, which here is realized as a scanning tunneling microscope tip, realizes the bottom arm of the beam splitter in Fig.~\ref{interferometer_scheme_nodot} and makes contact with both helical states on both edges, represented as blue and red lines.
The helical edges represent the other two arms of the beam splitter.
Additionally, molecular junctions have also been suggested as a viable option for realizing ballistic Andreev interferometers~\cite{Plaszko2020}.

\section{Conclusions}
\label{conc}
Electron coolers based on normal metal–insulator–su-perconductor (NIS) are considered highly promising because they enable refrigeration directly on-chip and controlled by electric means.

In this paper, we propose a possible improvement of such a class of refrigerators which makes use of an Andreev interferometer. We show that this approach yields a cooler with enhanced performance, specifically with a cooling power per channel that can be up to five orders of magnitude larger than that of a traditional NIS cooler.
This result originates from
the destructive interference occurring when the superconducting phase difference is equal to $\pi$.
Such interference leads to a significantly suppressed Andreev reflection below the superconducting gap, reducing Joule heating while still enabling substantial electron extraction above the gap.
Remarkably, we demonstrate that the minimal electronic temperature achievable is close to the lower bound set by the maximum heat current that can be extracted from a terminal in a two-terminal system.
In addition, we derive an approximate expression for this bound, which is valid in the regime of low bath temperatures.
Optimal cooling is achieved when transport in the beam splitter is ballistic and when the coupling to the two superconducting regions is symmetric. To address this, we analyze a beam splitter with arms of different lengths and with asymmetric arms and calculate the optimal cooling power.
At higher temperatures, this is found to exhibit relatively low sensitivity to small length differences and degree of asymmetry.
On the other hand, the minimal temperature achievable
is highly sensitive due to maximal refrigeration occurring at lower temperatures.

Finally, we propose various potential implementations of a ballistic Andreev interferometer cooler, focusing on materials such as semiconducting heterostructures, graphene, and topological insulators.

\begin{acknowledgments}
We thank Rosario Fazio and
Francesco Giazotto for the fruitful discussions.
FT acknowledges funding from MUR-PRIN 2022 - Grant No. 2022B9P8LN
- (PE3)-Project NEThEQS “Non-equilibrium coherent
thermal effects in quantum systems” in PNRR Mission 4 -
Component 2 - Investment 1.1 “Fondo per il Programma
Nazionale di Ricerca e Progetti di Rilevante Interesse
Nazionale (PRIN)” funded by the European Union - Next
Generation EU and the PNRR MUR project
PE0000023-NQSTI, and from the Royal Society through the International Exchanges between the UK and Italy (Grants No.
IEC R2 192166).\\
The software used to produce the plots in this article is openly available at Zenodo \cite{myplottingtool2025}.

\end{acknowledgments}

\appendix

\section{Calculation of the Andreev reflection for a beam splitter with arms of different
length}
\label{App-diffL}

By composing the scattering matrices of the beam splitter and of the two NS interfaces, as detailed in Sec.~\ref{Model}, we can compute the Andreev reflection amplitude ($S^{-+}$) and the normal reflection amplitude ($S^{++}$)  for the AI. They are given by
\begin{equation}
\label{sss}
    \begin{split}
        &S^{-+}(E) = \frac{2\alpha (1 + e^{i\Delta\phi})(1 - \alpha^2)}{4e^{i\Delta\phi} - \alpha^2 (1+e^{i\Delta\phi})^2}, \\
        &S^{++}(E) = \frac{(1 - \cos \Delta\phi) \alpha^2}{-2 + \alpha^2 (1 + \cos \Delta\phi)},
    \end{split}
\end{equation}
where $\alpha(E,T) = \exp\{-i\arccos[E/|\Delta(T)|]\}$ depends on both energy $E$ and temperature $T$.
Notice that the scattering coefficients are defined as $R^{\alpha\beta}(E)=|S^{\alpha\beta}(E)|^2$.
For $\Delta\phi = 0$, Eqs.~(\ref{sss}) reduce to
\begin{equation}
\label{fi0}
    \begin{split}
        &S^{-+}(E) = \alpha, \\
        &S^{++}(E) = 0,
    \end{split}
\end{equation}
which is simply the result for a single ideal NS junction. For $\Delta\phi = \pi$, however, we have
\begin{equation}
\label{fipi}
    \begin{split}
        &S^{-+}(E) = 0, \\
        &S^{++}(E) = -\alpha^2,
    \end{split}
\end{equation}
i.e.~Andreev reflection vanishes as a result of destructive interference.
It is interesting to notice that, above the gap ($E>\Delta$), normal reflection for $\Delta\phi = \pi$  decays faster than Andreev reflection for $\Delta\phi = 0$.
It is crucial to emphasize that the results presented in Eqs.~(\ref{fi0}) and (\ref{fipi}) are significantly influenced by the symmetry of the beam splitter.

Here, we aim to investigate the impact of an asymmetric beam splitter by analyzing a configuration where the two arms have different lengths.
To achieve this, we introduce the dynamical phase $\exp[ik(E)L]$ acquired by electrons traveling a distance $L$.
The wavevector $k(E)$ is defined as
\begin{equation*}
    k(E) = k_F \sqrt{1 + \frac{E}{\epsilon_F}},
\end{equation*}
where $\epsilon_F$ is the Fermi energy, $k_F=\sqrt{2m\epsilon_F / \hbar^2}$ is the Fermi wavevector, and $m$ the electron mass.
The overall scattering amplitudes, used to produce the plots in Fig.~\ref{deltaTNvsV_Tph0.2Al_vsdL}, are computed by including the dynamical phases relative to the two arms of lengths $L_1$ and $L_2$ into the composition procedure.

In the following we derive an approximate expression for the Andreev reflection amplitude valid for small differences in length, $\delta L = L_2 - L_1$, and for energies $E$ that are not significantly larger than $\Delta$, when $\Delta \ll \epsilon_F$.
Under the latter condition, the wavevector can be approximated as
\begin{equation*}
    k(E) \simeq k_F + \frac{E}{\hbar v_F},
\end{equation*}
where $v_F= \sqrt{2\epsilon_F/m}$ is the Fermi velocity, and the phase acquired across arm 1 (2), with length $L_{1(2)}$,  is given by
\begin{equation}
    \chi_{1(2)}(E) = k(E)L_{1(2)} = \left ( \frac{2\pi\xi}{\lambda_F} + \frac{E}{\Delta} \right) \frac{L_{1(2)}}{\xi}.
\end{equation}
In the above expression, $\xi=\hbar v_F/ \Delta$ is the coherence length of the superconductor and $\lambda_F=2\pi/k_F$ is the Fermi wavelength. 
For materials such as aluminum and InAs one finds~\cite{Blasi2023} that $\xi \sim 1\mu \mathrm{m}$, $\xi/\lambda_F \sim 10$ and for a realistic system~\cite{Blasi2023} we can take $L_1$ and $L_2$ of the order of $\xi$.
At $\Delta\phi = \pi$, we obtain the expression
\begin{widetext}
    \begin{equation}
        S^{-+}(E) = \frac{2 \alpha (e^{2i L_1 E/(\hbar v_F)} - e^{2i L_2 E/(\hbar v_F)}) [1 + \alpha^2 e^{2i(L_1+L_2)E/(\hbar v_F)}]}{-4 + \alpha^2 (e^{2i L_1 E/(\hbar v_F)} - e^{2i L_2 E/(\hbar v_F)})^2}.
    \end{equation}
\end{widetext}
This expression is generally non-zero.
By expanding the exponential terms for small $\delta L{/\xi}$, we can approximate $S^{-+}(E)$ as follows
\begin{equation}
    S^{-+}(E) \simeq - i \alpha \frac{E}{\Delta}\frac{\delta L}{ \xi} [1 + \alpha^2 e^{4i L_1 E/(\hbar v_F)}] .
\end{equation}
This approximation indicates that $S^{-+}(E)$ scales as $\delta L/\xi$, so that $R^{-+}(E)$ scales with $(\delta L/\xi)^2$.

We finally notice that the scattering amplitudes obtained with this approximation, for example with the parameters used in Fig.~\ref{deltaTNvsV_Tph0.2Al_vsdL}, are in very good agreement with the exact ones.

\section{Derivation of Eq.~(\ref{whitney_approx})}
\label{App-7}

The energy balance equation, Eq.~(\ref{baleq}), for the maximum cooling power given by Eq.~(\ref{whitney_bound}), reads
\begin{equation}
\label{tmin}
    \frac{N \pi^2 (k_\mathrm{B}T_{\rm N,min})^2}{12h} + \Sigma \mathcal{V}(T_{\rm N,min}^5 - T_\mathrm{ph}^5) = 0.
\end{equation}
For a fixed bath temperature $T_\mathrm{ph}$, this equation gives the minimal temperature achievable $T_{\rm N,min}$.
The above equation can be written as
\begin{equation}
\label{ttilde}
     \tilde{T}_{\rm N,min}^2 = \frac{12 h \Sigma \mathcal{V} T_\mathrm{C}^3}{N \pi^2 k_\mathrm{B}^2} (\tilde{T}_\mathrm{ph}^5 - \tilde{T}_{\rm N,min}^5),
\end{equation}
where the temperatures are measured in units of the critical temperature $T_\mathrm{C}$, i.e.~$\tilde T_{\rm N,min}=T_{\rm N,min}/T_\mathrm{C}$ and $\tilde T_{\rm ph}=T_{\rm ph}/T_\mathrm{C}$.
Since the second term on the r.h.s. of Eq.~(\ref{ttilde}) is negative, the l.h.s. can be upper bounded as
\begin{equation}
    \tilde{T}_{\rm N,min} \leq \sqrt{\frac{12 h \Sigma}{\pi^2 k_\mathrm{B}^2} \frac{ \mathcal{V} T_\mathrm{C}^3}{N} \tilde{T}_\mathrm{ph}^5}.
\end{equation}
Notice that Eq.~(\ref{tmin}) can also be written as
\begin{equation}
    \tilde{T}_\mathrm{N,min}^2 =
    \frac{\tilde{T}_\mathrm{ph}^5}{ 
     \frac{\pi^2 k_\mathrm{B}^2}{12 h \Sigma} \frac{N}{ \mathcal{V} T_\mathrm{C}^3} +\tilde{T}_\mathrm{N,min}^3}  ,
\end{equation}
which can be lower bounded as
\begin{equation}
\label{low}
    \tilde{T}_\mathrm{N} \geq \sqrt{\frac{\tilde{T}_\mathrm{ph}^5}{\frac{\pi^2 k_\mathrm{B}^2}{12 h \Sigma} \frac{N}{ \mathcal{V} T_\mathrm{C}^3} + \tilde{T}_\mathrm{ph}^3}},
\end{equation}
when $\tilde T_\mathrm{N,min}<\tilde T_\mathrm{ph}$.
It turns out that lower and upper bounds coincide for small enough $\tilde T_\mathrm{ph}$, as one can expect by neglecting the second term in the denominator of Eq.~(\ref{low}).
This shows that Eq.~(\ref{whitney_approx}) represents a good approximation of the minimal temperature achievable for small bath temperatures, while the lower bound is approached for higher values of the bath temperature.\\

\section{Andreev reflection in the multichannel case}
\label{App-multi}

In this Appendix, we show that in an AI with multi-channel leads, destructive interference occurs for $\Delta\phi=\pi$, resulting in a vanishing Andreev reflection probability, provided the beam splitter arms are symmetric.
For multichannel leads, each with $N$ open channels, the scattering matrix of the beam splitter with symmetric arms, as given in Eq.~(\ref{smat}) for a single channel, becomes
\begin{equation}
\label{sbs}
    S_\mathrm{BS} = \begin{pmatrix}
        \mathbf{r} & \mathbf{t'} & \mathbf{t'} \\ \mathbf{t} & \mathbf{a} & \mathbf{b} \\ \mathbf{t} & \mathbf{b} & \mathbf{a}
    \end{pmatrix},
\end{equation}
where $\mathbf{r}$ represents the reflection matrix for lead N, $\mathbf{t}$ is the transmission matrix from lead N to lead 1 and to lead 2, and $\mathbf{t'}$ is the transmission matrix from lead 1 and lead 2 to lead N. 
The symmetry of the beam splitter is reflected in the fact that the transmission from lead N to leads 1 and 2 is described by the same matrix $\mathbf{t}$, and similarly for the matrix $\mathbf{t'}$. Each matrix, denoted by a bold letter and potentially energy-dependent, has dimension $N\times N$.

The matrix representing both ideal NS interfaces for the case $\Delta \phi = \pi$ can be written as
\begin{equation}
    S_\mathrm{NS} = \begin{pmatrix}
        \bm{\alpha} & 0 \\ 0 & -\bm{\alpha}
    \end{pmatrix},
    \label{sns}
\end{equation}
where the diagonal blocks correspond to the two arms of the junction. The matrix $\bm{\alpha}$, with dimension $N\times N$, is proportional to the identity matrix and contains the Andreev reflection amplitudes for an ideal NS junction, given by $(\bm{\alpha})_{ij} =\delta_{ij} \exp\{-i\arccos[E/\Delta(T)]\}$, as shown in Eqs.~(\ref{rhe}) and (\ref{reh}). The minus sign in the bottom-right block accounts for the phase difference $\Delta \phi = \pi$.

The total Andreev reflection matrix $\mathbf{S^{-+}}$, with dimension $N\times N$, can be obtained by composing~\cite{Datta1997} the scattering matrices in Eqs.~(\ref{sbs}) and (\ref{sns}), resulting in
\begin{widetext}
    \begin{equation}
        \mathbf{S}^{-+} = \begin{matrix} \begin{pmatrix} \mathbf{t'}_\mathrm{h} & \mathbf{t'}_\mathrm{h} \end{pmatrix} \\\mbox{} \end{matrix} \begin{pmatrix}
        \bm{\alpha} & 0 \\ 0 & -\bm{\alpha}
    \end{pmatrix} \left [ \mathbb{I} - \begin{pmatrix} \mathbf{a} & \mathbf{b} \\ \mathbf{b} & \mathbf{a} \end{pmatrix}  \begin{pmatrix}  \bm{\alpha} & 0 \\ 0 & -\bm{\alpha} \end{pmatrix} \begin{pmatrix} \mathbf{a}_\mathrm{h} & \mathbf{b}_\mathrm{h} \\ \mathbf{b}_\mathrm{h} & \mathbf{a}_\mathrm{h} \end{pmatrix} \begin{pmatrix}  \bm{\alpha} & 0 \\ 0 & -\bm{\alpha} \end{pmatrix}  \right]^{-1} \begin{pmatrix} \mathbf{t} \\ \mathbf{t} \end{pmatrix},
    \label{smp}
    \end{equation}
\end{widetext}
where the subscript h indicates the scattering matrix for the hole degree of freedom, which can be derived from matrix (\ref{sbs}) using particle-hole symmetry~\cite{Lambert1998}.
At the lowest order, i.e.~neglecting multiple reflections in the beam splitter, $\mathbf{S}^{-+}$ can be approximated as
\begin{equation}
        \mathbf{S}^{-+} \simeq \begin{matrix} \begin{pmatrix} \mathbf{t'}_\mathrm{h} & \mathbf{t'}_\mathrm{h} \end{pmatrix} \\\mbox{} \end{matrix} \begin{pmatrix}
        \bm{\alpha} & 0 \\ 0 & -\bm{\alpha}
    \end{pmatrix}
\begin{pmatrix} \mathbf{t} \\ \mathbf{t} \end{pmatrix},
\end{equation}
which equals zero due to the equality of the components of both vectors, indicating the symmetry of the beam splitter arms, and the structure of the matrix $S_\mathrm{NS}$ in Eq.~(\ref{sns}), where the diagonal blocks differ only by a minus sign.
By explicitly calculating all matrix products in Eq.~(\ref{smp}) and using similar arguments it is possible to show that even the total Andreev reflection matrix (which accounts for all multiple reflections) vanishes identically, i.e. $\mathbf{S}^{-+}=0$ when $\Delta\phi=\pi$.

This shows that the Andreev reflection vanishes when $\Delta\phi=\pi$ even in the multichannel case as long as the beam splitter is symmetric.
Note that the matrices $\mathbf{t}$ and $\mathbf{t_h'}$ can be energy-dependent and need not be diagonal, allowing for mixing between channels, provided the mixing is symmetric between the two arms.

\bibliography{mybib}

\begin{thebibliography}{84}%
\makeatletter
\providecommand \@ifxundefined [1]{%
 \@ifx{#1\undefined}
}%
\providecommand \@ifnum [1]{%
 \ifnum #1\expandafter \@firstoftwo
 \else \expandafter \@secondoftwo
 \fi
}%
\providecommand \@ifx [1]{%
 \ifx #1\expandafter \@firstoftwo
 \else \expandafter \@secondoftwo
 \fi
}%
\providecommand \natexlab [1]{#1}%
\providecommand \enquote  [1]{``#1''}%
\providecommand \bibnamefont  [1]{#1}%
\providecommand \bibfnamefont [1]{#1}%
\providecommand \citenamefont [1]{#1}%
\providecommand \href@noop [0]{\@secondoftwo}%
\providecommand \href [0]{\begingroup \@sanitize@url \@href}%
\providecommand \@href[1]{\@@startlink{#1}\@@href}%
\providecommand \@@href[1]{\endgroup#1\@@endlink}%
\providecommand \@sanitize@url [0]{\catcode `\\12\catcode `\$12\catcode `\&12\catcode `\#12\catcode `\^12\catcode `\_12\catcode `\%12\relax}%
\providecommand \@@startlink[1]{}%
\providecommand \@@endlink[0]{}%
\providecommand \url  [0]{\begingroup\@sanitize@url \@url }%
\providecommand \@url [1]{\endgroup\@href {#1}{\urlprefix }}%
\providecommand \urlprefix  [0]{URL }%
\providecommand \Eprint [0]{\href }%
\providecommand \doibase [0]{https://doi.org/}%
\providecommand \selectlanguage [0]{\@gobble}%
\providecommand \bibinfo  [0]{\@secondoftwo}%
\providecommand \bibfield  [0]{\@secondoftwo}%
\providecommand \translation [1]{[#1]}%
\providecommand \BibitemOpen [0]{}%
\providecommand \bibitemStop [0]{}%
\providecommand \bibitemNoStop [0]{.\EOS\space}%
\providecommand \EOS [0]{\spacefactor3000\relax}%
\providecommand \BibitemShut  [1]{\csname bibitem#1\endcsname}%
\let\auto@bib@innerbib\@empty
\bibitem [{\citenamefont {Giazotto}\ \emph {et~al.}(2006{\natexlab{a}})\citenamefont {Giazotto}, \citenamefont {Heikkil\"a}, \citenamefont {Luukanen}, \citenamefont {Savin},\ and\ \citenamefont {Pekola}}]{Giazotto2006}%
  \BibitemOpen
  \bibfield  {author} {\bibinfo {author} {\bibfnamefont {F.}~\bibnamefont {Giazotto}}, \bibinfo {author} {\bibfnamefont {T.~T.}\ \bibnamefont {Heikkil\"a}}, \bibinfo {author} {\bibfnamefont {A.}~\bibnamefont {Luukanen}}, \bibinfo {author} {\bibfnamefont {A.~M.}\ \bibnamefont {Savin}},\ and\ \bibinfo {author} {\bibfnamefont {J.~P.}\ \bibnamefont {Pekola}},\ }\bibfield  {title} {\bibinfo {title} {Opportunities for mesoscopics in thermometry and refrigeration: Physics and applications},\ }\href {https://doi.org/10.1103/RevModPhys.78.217} {\bibfield  {journal} {\bibinfo  {journal} {Rev. Mod. Phys.}\ }\textbf {\bibinfo {volume} {78}},\ \bibinfo {pages} {217} (\bibinfo {year} {2006}{\natexlab{a}})}\BibitemShut {NoStop}%
\bibitem [{\citenamefont {Benenti}\ \emph {et~al.}(2017)\citenamefont {Benenti}, \citenamefont {Casati}, \citenamefont {Saito},\ and\ \citenamefont {Whitney}}]{Benenti2017}%
  \BibitemOpen
  \bibfield  {author} {\bibinfo {author} {\bibfnamefont {G.}~\bibnamefont {Benenti}}, \bibinfo {author} {\bibfnamefont {G.}~\bibnamefont {Casati}}, \bibinfo {author} {\bibfnamefont {K.}~\bibnamefont {Saito}},\ and\ \bibinfo {author} {\bibfnamefont {R.}~\bibnamefont {Whitney}},\ }\bibfield  {title} {\bibinfo {title} {Fundamental aspects of steady-state conversion of heat to work at the nanoscale},\ }\href {https://doi.org/https://doi.org/10.1016/j.physrep.2017.05.008} {\bibfield  {journal} {\bibinfo  {journal} {Phys. Rep.}\ }\textbf {\bibinfo {volume} {694}},\ \bibinfo {pages} {1} (\bibinfo {year} {2017})}\BibitemShut {NoStop}%
\bibitem [{\citenamefont {Pekola}\ and\ \citenamefont {Karimi}(2021)}]{Pekola2021}%
  \BibitemOpen
  \bibfield  {author} {\bibinfo {author} {\bibfnamefont {J.~P.}\ \bibnamefont {Pekola}}\ and\ \bibinfo {author} {\bibfnamefont {B.}~\bibnamefont {Karimi}},\ }\bibfield  {title} {\bibinfo {title} {Colloquium: Quantum heat transport in condensed matter systems},\ }\href {https://doi.org/10.1103/RevModPhys.93.041001} {\bibfield  {journal} {\bibinfo  {journal} {Rev. Mod. Phys.}\ }\textbf {\bibinfo {volume} {93}},\ \bibinfo {pages} {041001} (\bibinfo {year} {2021})}\BibitemShut {NoStop}%
\bibitem [{\citenamefont {Majidi}\ \emph {et~al.}(2024)\citenamefont {Majidi}, \citenamefont {Bergfield}, \citenamefont {Maisi}, \citenamefont {Höfer}, \citenamefont {Courtois},\ and\ \citenamefont {Winkelmann}}]{Majidi2024}%
  \BibitemOpen
  \bibfield  {author} {\bibinfo {author} {\bibfnamefont {D.}~\bibnamefont {Majidi}}, \bibinfo {author} {\bibfnamefont {J.~P.}\ \bibnamefont {Bergfield}}, \bibinfo {author} {\bibfnamefont {V.}~\bibnamefont {Maisi}}, \bibinfo {author} {\bibfnamefont {J.}~\bibnamefont {Höfer}}, \bibinfo {author} {\bibfnamefont {H.}~\bibnamefont {Courtois}},\ and\ \bibinfo {author} {\bibfnamefont {C.~B.}\ \bibnamefont {Winkelmann}},\ }\bibfield  {title} {\bibinfo {title} {{Heat transport at the nanoscale and ultralow temperatures—Implications for quantum technologies}},\ }\href {https://doi.org/10.1063/5.0204207} {\bibfield  {journal} {\bibinfo  {journal} {App. Phys. Lett.}\ }\textbf {\bibinfo {volume} {124}},\ \bibinfo {pages} {140504} (\bibinfo {year} {2024})}\BibitemShut {NoStop}%
\bibitem [{\citenamefont {Haack}\ and\ \citenamefont {Giazotto}(2019)}]{Haack2019}%
  \BibitemOpen
  \bibfield  {author} {\bibinfo {author} {\bibfnamefont {G.}~\bibnamefont {Haack}}\ and\ \bibinfo {author} {\bibfnamefont {F.}~\bibnamefont {Giazotto}},\ }\bibfield  {title} {\bibinfo {title} {{Efficient and tunable Aharonov-Bohm quantum heat engine}},\ }\href {https://doi.org/10.1103/PhysRevB.100.235442} {\bibfield  {journal} {\bibinfo  {journal} {Phys. Rev. B}\ }\textbf {\bibinfo {volume} {100}},\ \bibinfo {pages} {235442} (\bibinfo {year} {2019})}\BibitemShut {NoStop}%
\bibitem [{\citenamefont {Haack}\ and\ \citenamefont {Giazotto}(2021)}]{Haack2021}%
  \BibitemOpen
  \bibfield  {author} {\bibinfo {author} {\bibfnamefont {G.}~\bibnamefont {Haack}}\ and\ \bibinfo {author} {\bibfnamefont {F.}~\bibnamefont {Giazotto}},\ }\bibfield  {title} {\bibinfo {title} {{Nonlinear regime for enhanced performance of an Aharonov–Bohm heat engine}},\ }\href {https://doi.org/10.1116/5.0064936} {\bibfield  {journal} {\bibinfo  {journal} {AVS Quantum Science}\ }\textbf {\bibinfo {volume} {3}},\ \bibinfo {pages} {046801} (\bibinfo {year} {2021})}\BibitemShut {NoStop}%
\bibitem [{\citenamefont {Giazotto}\ and\ \citenamefont {Mart{\'\i}nez-P{\'e}rez}(2012)}]{Giazotto2012}%
  \BibitemOpen
  \bibfield  {author} {\bibinfo {author} {\bibfnamefont {F.}~\bibnamefont {Giazotto}}\ and\ \bibinfo {author} {\bibfnamefont {M.~J.}\ \bibnamefont {Mart{\'\i}nez-P{\'e}rez}},\ }\bibfield  {title} {\bibinfo {title} {{The Josephson heat interferometer}},\ }\href {https://doi.org/10.1038/nature11702} {\bibfield  {journal} {\bibinfo  {journal} {Nature}\ }\textbf {\bibinfo {volume} {492}},\ \bibinfo {pages} {401} (\bibinfo {year} {2012})}\BibitemShut {NoStop}%
\bibitem [{\citenamefont {Mart{\'\i}nez-P{\'e}rez}\ and\ \citenamefont {Giazotto}(2013)}]{Martinez2013}%
  \BibitemOpen
  \bibfield  {author} {\bibinfo {author} {\bibfnamefont {M.~J.}\ \bibnamefont {Mart{\'\i}nez-P{\'e}rez}}\ and\ \bibinfo {author} {\bibfnamefont {F.}~\bibnamefont {Giazotto}},\ }\bibfield  {title} {\bibinfo {title} {{Efficient phase-tunable Josephson thermal rectifier}},\ }\href {https://doi.org/10.1063/1.4804550} {\bibfield  {journal} {\bibinfo  {journal} {App. Phys. Lett.}\ }\textbf {\bibinfo {volume} {102}},\ \bibinfo {pages} {182602} (\bibinfo {year} {2013})}\BibitemShut {NoStop}%
\bibitem [{\citenamefont {Bosisio}\ \emph {et~al.}(2015)\citenamefont {Bosisio}, \citenamefont {Valentini}, \citenamefont {Mazza}, \citenamefont {Benenti}, \citenamefont {Fazio}, \citenamefont {Giovannetti},\ and\ \citenamefont {Taddei}}]{Bosisio2015}%
  \BibitemOpen
  \bibfield  {author} {\bibinfo {author} {\bibfnamefont {R.}~\bibnamefont {Bosisio}}, \bibinfo {author} {\bibfnamefont {S.}~\bibnamefont {Valentini}}, \bibinfo {author} {\bibfnamefont {F.}~\bibnamefont {Mazza}}, \bibinfo {author} {\bibfnamefont {G.}~\bibnamefont {Benenti}}, \bibinfo {author} {\bibfnamefont {R.}~\bibnamefont {Fazio}}, \bibinfo {author} {\bibfnamefont {V.}~\bibnamefont {Giovannetti}},\ and\ \bibinfo {author} {\bibfnamefont {F.}~\bibnamefont {Taddei}},\ }\bibfield  {title} {\bibinfo {title} {Magnetic thermal switch for heat management at the nanoscale},\ }\href {https://doi.org/10.1103/PhysRevB.91.205420} {\bibfield  {journal} {\bibinfo  {journal} {Phys. Rev. B}\ }\textbf {\bibinfo {volume} {91}},\ \bibinfo {pages} {205420} (\bibinfo {year} {2015})}\BibitemShut {NoStop}%
\bibitem [{\citenamefont {Fornieri}\ \emph {et~al.}(2016)\citenamefont {Fornieri}, \citenamefont {Blanc}, \citenamefont {Bosisio}, \citenamefont {D'Ambrosio},\ and\ \citenamefont {Giazotto}}]{Fornieri2016}%
  \BibitemOpen
  \bibfield  {author} {\bibinfo {author} {\bibfnamefont {A.}~\bibnamefont {Fornieri}}, \bibinfo {author} {\bibfnamefont {C.}~\bibnamefont {Blanc}}, \bibinfo {author} {\bibfnamefont {R.}~\bibnamefont {Bosisio}}, \bibinfo {author} {\bibfnamefont {S.}~\bibnamefont {D'Ambrosio}},\ and\ \bibinfo {author} {\bibfnamefont {F.}~\bibnamefont {Giazotto}},\ }\bibfield  {title} {\bibinfo {title} {{Nanoscale phase engineering of thermal transport with a Josephson heat modulator}},\ }\href {https://doi.org/10.1038/nnano.2015.281} {\bibfield  {journal} {\bibinfo  {journal} {Nat. Nanotechnol.}\ }\textbf {\bibinfo {volume} {11}},\ \bibinfo {pages} {258} (\bibinfo {year} {2016})}\BibitemShut {NoStop}%
\bibitem [{\citenamefont {Solinas}\ \emph {et~al.}(2016)\citenamefont {Solinas}, \citenamefont {Bosisio},\ and\ \citenamefont {Giazotto}}]{Solinas2016}%
  \BibitemOpen
  \bibfield  {author} {\bibinfo {author} {\bibfnamefont {P.}~\bibnamefont {Solinas}}, \bibinfo {author} {\bibfnamefont {R.}~\bibnamefont {Bosisio}},\ and\ \bibinfo {author} {\bibfnamefont {F.}~\bibnamefont {Giazotto}},\ }\bibfield  {title} {\bibinfo {title} {Microwave quantum refrigeration based on the josephson effect},\ }\href {https://doi.org/10.1103/PhysRevB.93.224521} {\bibfield  {journal} {\bibinfo  {journal} {Phys. Rev. B}\ }\textbf {\bibinfo {volume} {93}},\ \bibinfo {pages} {224521} (\bibinfo {year} {2016})}\BibitemShut {NoStop}%
\bibitem [{\citenamefont {Fornieri}\ \emph {et~al.}(2017)\citenamefont {Fornieri}, \citenamefont {Timossi}, \citenamefont {Virtanen}, \citenamefont {Solinas},\ and\ \citenamefont {Giazotto}}]{Fornieri2017}%
  \BibitemOpen
  \bibfield  {author} {\bibinfo {author} {\bibfnamefont {A.}~\bibnamefont {Fornieri}}, \bibinfo {author} {\bibfnamefont {G.}~\bibnamefont {Timossi}}, \bibinfo {author} {\bibfnamefont {P.}~\bibnamefont {Virtanen}}, \bibinfo {author} {\bibfnamefont {P.}~\bibnamefont {Solinas}},\ and\ \bibinfo {author} {\bibfnamefont {F.}~\bibnamefont {Giazotto}},\ }\bibfield  {title} {\bibinfo {title} {{0--$\pi$ phase-controllable thermal Josephson junction}},\ }\href {https://doi.org/10.1038/nnano.2017.25} {\bibfield  {journal} {\bibinfo  {journal} {Nat. Nanotechnol.}\ }\textbf {\bibinfo {volume} {12}},\ \bibinfo {pages} {425} (\bibinfo {year} {2017})}\BibitemShut {NoStop}%
\bibitem [{\citenamefont {Sothmann}\ \emph {et~al.}(2017)\citenamefont {Sothmann}, \citenamefont {Giazotto},\ and\ \citenamefont {Hankiewicz}}]{Sothmann2017}%
  \BibitemOpen
  \bibfield  {author} {\bibinfo {author} {\bibfnamefont {B.}~\bibnamefont {Sothmann}}, \bibinfo {author} {\bibfnamefont {F.}~\bibnamefont {Giazotto}},\ and\ \bibinfo {author} {\bibfnamefont {E.~M.}\ \bibnamefont {Hankiewicz}},\ }\bibfield  {title} {\bibinfo {title} {{High-efficiency thermal switch based on topological Josephson junctions}},\ }\href {https://doi.org/10.1088/1367-2630/aa60d4} {\bibfield  {journal} {\bibinfo  {journal} {New J. Phys.}\ }\textbf {\bibinfo {volume} {19}},\ \bibinfo {pages} {023056} (\bibinfo {year} {2017})}\BibitemShut {NoStop}%
\bibitem [{\citenamefont {Hwang}\ \emph {et~al.}(2018)\citenamefont {Hwang}, \citenamefont {Giazotto},\ and\ \citenamefont {Sothmann}}]{Hwang2018}%
  \BibitemOpen
  \bibfield  {author} {\bibinfo {author} {\bibfnamefont {S.-Y.}\ \bibnamefont {Hwang}}, \bibinfo {author} {\bibfnamefont {F.}~\bibnamefont {Giazotto}},\ and\ \bibinfo {author} {\bibfnamefont {B.}~\bibnamefont {Sothmann}},\ }\bibfield  {title} {\bibinfo {title} {{Phase-coherent heat circulator based on multiterminal Josephson junctions}},\ }\href {https://doi.org/10.1103/PhysRevApplied.10.044062} {\bibfield  {journal} {\bibinfo  {journal} {Phys. Rev. Appl.}\ }\textbf {\bibinfo {volume} {10}},\ \bibinfo {pages} {044062} (\bibinfo {year} {2018})}\BibitemShut {NoStop}%
\bibitem [{\citenamefont {Acciai}\ \emph {et~al.}(2021)\citenamefont {Acciai}, \citenamefont {Hajiloo}, \citenamefont {Hassler},\ and\ \citenamefont {Splettstoesser}}]{Acciai2021}%
  \BibitemOpen
  \bibfield  {author} {\bibinfo {author} {\bibfnamefont {M.}~\bibnamefont {Acciai}}, \bibinfo {author} {\bibfnamefont {F.}~\bibnamefont {Hajiloo}}, \bibinfo {author} {\bibfnamefont {F.}~\bibnamefont {Hassler}},\ and\ \bibinfo {author} {\bibfnamefont {J.}~\bibnamefont {Splettstoesser}},\ }\bibfield  {title} {\bibinfo {title} {Phase-coherent heat circulators with normal or superconducting contacts},\ }\href {https://doi.org/10.1103/PhysRevB.103.085409} {\bibfield  {journal} {\bibinfo  {journal} {Phys. Rev. B}\ }\textbf {\bibinfo {volume} {103}},\ \bibinfo {pages} {085409} (\bibinfo {year} {2021})}\BibitemShut {NoStop}%
\bibitem [{\citenamefont {Huang}(2023)}]{Huang2023}%
  \BibitemOpen
  \bibfield  {author} {\bibinfo {author} {\bibfnamefont {C.-S.}\ \bibnamefont {Huang}},\ }\bibfield  {title} {\bibinfo {title} {{Phase dependence of thermal transport in graphene Josephson junctions}},\ }\href {https://doi.org/10.1103/PhysRevB.108.195433} {\bibfield  {journal} {\bibinfo  {journal} {Phys. Rev. B}\ }\textbf {\bibinfo {volume} {108}},\ \bibinfo {pages} {195433} (\bibinfo {year} {2023})}\BibitemShut {NoStop}%
\bibitem [{\citenamefont {Hwang}\ \emph {et~al.}(2024)\citenamefont {Hwang}, \citenamefont {Sothmann},\ and\ \citenamefont {L\'opez}}]{Hwang2024}%
  \BibitemOpen
  \bibfield  {author} {\bibinfo {author} {\bibfnamefont {S.-Y.}\ \bibnamefont {Hwang}}, \bibinfo {author} {\bibfnamefont {B.}~\bibnamefont {Sothmann}},\ and\ \bibinfo {author} {\bibfnamefont {R.}~\bibnamefont {L\'opez}},\ }\bibfield  {title} {\bibinfo {title} {{Phase-controlled heat modulation with Aharonov-Bohm interferometers}},\ }\href {https://doi.org/10.1103/PhysRevResearch.6.013215} {\bibfield  {journal} {\bibinfo  {journal} {Phys. Rev. Res.}\ }\textbf {\bibinfo {volume} {6}},\ \bibinfo {pages} {013215} (\bibinfo {year} {2024})}\BibitemShut {NoStop}%
\bibitem [{\citenamefont {Balduque}\ \emph {et~al.}(2024)\citenamefont {Balduque}, \citenamefont {Mecha},\ and\ \citenamefont {Sánchez}}]{Balduque2024}%
  \BibitemOpen
  \bibfield  {author} {\bibinfo {author} {\bibfnamefont {J.}~\bibnamefont {Balduque}}, \bibinfo {author} {\bibfnamefont {A.}~\bibnamefont {Mecha}},\ and\ \bibinfo {author} {\bibfnamefont {R.}~\bibnamefont {Sánchez}},\ }\bibfield  {title} {\bibinfo {title} {{Thermal junctions controlled with Aharonov–Bohm phases}},\ }\href {https://doi.org/10.1063/5.0218032} {\bibfield  {journal} {\bibinfo  {journal} {APL Quantum}\ }\textbf {\bibinfo {volume} {1}},\ \bibinfo {pages} {036120} (\bibinfo {year} {2024})}\BibitemShut {NoStop}%
\bibitem [{\citenamefont {Blasi}\ \emph {et~al.}(2023)\citenamefont {Blasi}, \citenamefont {Giazotto},\ and\ \citenamefont {Haack}}]{Blasi2023}%
  \BibitemOpen
  \bibfield  {author} {\bibinfo {author} {\bibfnamefont {G.}~\bibnamefont {Blasi}}, \bibinfo {author} {\bibfnamefont {F.}~\bibnamefont {Giazotto}},\ and\ \bibinfo {author} {\bibfnamefont {G.}~\bibnamefont {Haack}},\ }\bibfield  {title} {\bibinfo {title} {{Hybrid normal-superconducting Aharonov-Bohm quantum thermal device}},\ }\href {https://doi.org/10.1088/2058-9565/acacbf} {\bibfield  {journal} {\bibinfo  {journal} {Quantum Sci. Technol.}\ }\textbf {\bibinfo {volume} {8}},\ \bibinfo {pages} {015023} (\bibinfo {year} {2023})}\BibitemShut {NoStop}%
\bibitem [{\citenamefont {Muhonen}\ \emph {et~al.}(2012)\citenamefont {Muhonen}, \citenamefont {Meschke},\ and\ \citenamefont {Pekola}}]{Muhonen2012}%
  \BibitemOpen
  \bibfield  {author} {\bibinfo {author} {\bibfnamefont {J.~T.}\ \bibnamefont {Muhonen}}, \bibinfo {author} {\bibfnamefont {M.}~\bibnamefont {Meschke}},\ and\ \bibinfo {author} {\bibfnamefont {J.~P.}\ \bibnamefont {Pekola}},\ }\bibfield  {title} {\bibinfo {title} {Micrometre-scale refrigerators},\ }\href {https://doi.org/10.1088/0034-4885/75/4/046501} {\bibfield  {journal} {\bibinfo  {journal} {Rep. Prog. Phys.}\ }\textbf {\bibinfo {volume} {75}},\ \bibinfo {pages} {046501} (\bibinfo {year} {2012})}\BibitemShut {NoStop}%
\bibitem [{\citenamefont {Courtois}\ \emph {et~al.}(2014)\citenamefont {Courtois}, \citenamefont {Hekking}, \citenamefont {Nguyen},\ and\ \citenamefont {Winkelmann}}]{Courtois2014}%
  \BibitemOpen
  \bibfield  {author} {\bibinfo {author} {\bibfnamefont {H.}~\bibnamefont {Courtois}}, \bibinfo {author} {\bibfnamefont {F.}~\bibnamefont {Hekking}}, \bibinfo {author} {\bibfnamefont {H.}~\bibnamefont {Nguyen}},\ and\ \bibinfo {author} {\bibfnamefont {C.~B.}\ \bibnamefont {Winkelmann}},\ }\bibfield  {title} {\bibinfo {title} {Electronic coolers based on superconducting tunnel junctions: Fundamentals and applications},\ }\href {https://doi.org/10.1007/s10909-014-1101-0} {\bibfield  {journal} {\bibinfo  {journal} {J. Low Temp. Phys.}\ }\textbf {\bibinfo {volume} {91753}},\ \bibinfo {pages} {799} (\bibinfo {year} {2014})}\BibitemShut {NoStop}%
\bibitem [{\citenamefont {Ziabari}\ \emph {et~al.}(2016)\citenamefont {Ziabari}, \citenamefont {Zebarjadi}, \citenamefont {Vashaee},\ and\ \citenamefont {Shakouri}}]{Ziabari2016}%
  \BibitemOpen
  \bibfield  {author} {\bibinfo {author} {\bibfnamefont {A.}~\bibnamefont {Ziabari}}, \bibinfo {author} {\bibfnamefont {M.}~\bibnamefont {Zebarjadi}}, \bibinfo {author} {\bibfnamefont {D.}~\bibnamefont {Vashaee}},\ and\ \bibinfo {author} {\bibfnamefont {A.}~\bibnamefont {Shakouri}},\ }\bibfield  {title} {\bibinfo {title} {Nanoscale solid-state cooling: a review},\ }\href {https://doi.org/10.1088/0034-4885/79/9/095901} {\bibfield  {journal} {\bibinfo  {journal} {Rep. Prog. Phys.}\ }\textbf {\bibinfo {volume} {79}},\ \bibinfo {pages} {095901} (\bibinfo {year} {2016})}\BibitemShut {NoStop}%
\bibitem [{\citenamefont {Jones}\ \emph {et~al.}(2020)\citenamefont {Jones}, \citenamefont {Scheller}, \citenamefont {Prance}, \citenamefont {Kalyoncu}, \citenamefont {Zumb\"uhl},\ and\ \citenamefont {Haley}}]{Jones2020}%
  \BibitemOpen
  \bibfield  {author} {\bibinfo {author} {\bibfnamefont {A.}~\bibnamefont {Jones}}, \bibinfo {author} {\bibfnamefont {C.}~\bibnamefont {Scheller}}, \bibinfo {author} {\bibfnamefont {J.}~\bibnamefont {Prance}}, \bibinfo {author} {\bibfnamefont {Y.~B.}\ \bibnamefont {Kalyoncu}}, \bibinfo {author} {\bibfnamefont {D.~M.}\ \bibnamefont {Zumb\"uhl}},\ and\ \bibinfo {author} {\bibfnamefont {R.~P.}\ \bibnamefont {Haley}},\ }\bibfield  {title} {\bibinfo {title} {Progress in cooling nanoelectronic devices to ultra-low temperatures},\ }\href {https://doi.org/10.1007/s10909-020-02472-9} {\bibfield  {journal} {\bibinfo  {journal} {J. Low Temp. Phys.}\ }\textbf {\bibinfo {volume} {201}},\ \bibinfo {pages} {772} (\bibinfo {year} {2020})}\BibitemShut {NoStop}%
\bibitem [{\citenamefont {Cao}(2021)}]{Cao2021}%
  \BibitemOpen
  \bibfield  {author} {\bibinfo {author} {\bibfnamefont {H.}~\bibnamefont {Cao}},\ }\bibfield  {title} {\bibinfo {title} {{Refrigeration below 1 Kelvin}},\ }\href {https://doi.org/10.1007/s10909-021-02606-7} {\bibfield  {journal} {\bibinfo  {journal} {J. Low Temp. Phys.}\ }\textbf {\bibinfo {volume} {204}},\ \bibinfo {pages} {175} (\bibinfo {year} {2021})}\BibitemShut {NoStop}%
\bibitem [{\citenamefont {Nahum}\ \emph {et~al.}(1994)\citenamefont {Nahum}, \citenamefont {Eiles},\ and\ \citenamefont {Martinis}}]{Nahum1994}%
  \BibitemOpen
  \bibfield  {author} {\bibinfo {author} {\bibfnamefont {M.}~\bibnamefont {Nahum}}, \bibinfo {author} {\bibfnamefont {T.~M.}\ \bibnamefont {Eiles}},\ and\ \bibinfo {author} {\bibfnamefont {J.~M.}\ \bibnamefont {Martinis}},\ }\bibfield  {title} {\bibinfo {title} {Electronic microrefrigerator based on a normal-insulator-superconductor tunnel junction},\ }\href {https://doi.org/10.1063/1.112456} {\bibfield  {journal} {\bibinfo  {journal} {App. Phys. Lett.}\ }\textbf {\bibinfo {volume} {65}},\ \bibinfo {pages} {3123} (\bibinfo {year} {1994})}\BibitemShut {NoStop}%
\bibitem [{\citenamefont {Bardas}\ and\ \citenamefont {Averin}(1995)}]{Bardas1995}%
  \BibitemOpen
  \bibfield  {author} {\bibinfo {author} {\bibfnamefont {A.}~\bibnamefont {Bardas}}\ and\ \bibinfo {author} {\bibfnamefont {D.}~\bibnamefont {Averin}},\ }\bibfield  {title} {\bibinfo {title} {Peltier effect in normal-metal--superconductor microcontacts},\ }\href {https://doi.org/10.1103/PhysRevB.52.12873} {\bibfield  {journal} {\bibinfo  {journal} {Phys. Rev. B}\ }\textbf {\bibinfo {volume} {52}},\ \bibinfo {pages} {12873} (\bibinfo {year} {1995})}\BibitemShut {NoStop}%
\bibitem [{\citenamefont {Leivo}\ \emph {et~al.}(1996)\citenamefont {Leivo}, \citenamefont {Pekola},\ and\ \citenamefont {Averin}}]{Leivo1996}%
  \BibitemOpen
  \bibfield  {author} {\bibinfo {author} {\bibfnamefont {M.}~\bibnamefont {Leivo}}, \bibinfo {author} {\bibfnamefont {J.}~\bibnamefont {Pekola}},\ and\ \bibinfo {author} {\bibfnamefont {D.}~\bibnamefont {Averin}},\ }\bibfield  {title} {\bibinfo {title} {{Efficient Peltier refrigeration by a pair of normal metal/insulator/superconductor junctions}},\ }\href {https://doi.org/10.1063/1.115651} {\bibfield  {journal} {\bibinfo  {journal} {App. Phys. Lett.}\ }\textbf {\bibinfo {volume} {68}},\ \bibinfo {pages} {1996} (\bibinfo {year} {1996})}\BibitemShut {NoStop}%
\bibitem [{\citenamefont {Fisher}\ \emph {et~al.}(1999)\citenamefont {Fisher}, \citenamefont {Ullom},\ and\ \citenamefont {Nahum}}]{Fisher1999}%
  \BibitemOpen
  \bibfield  {author} {\bibinfo {author} {\bibfnamefont {P.~A.}\ \bibnamefont {Fisher}}, \bibinfo {author} {\bibfnamefont {J.~N.}\ \bibnamefont {Ullom}},\ and\ \bibinfo {author} {\bibfnamefont {M.}~\bibnamefont {Nahum}},\ }\bibfield  {title} {\bibinfo {title} {High-power on-chip microrefrigerator based on a normal- metal/insulator/superconductor tunnel junction},\ }\href {https://doi.org/10.1063/1.123943} {\bibfield  {journal} {\bibinfo  {journal} {Appl. Phys. Lett.}\ }\textbf {\bibinfo {volume} {74}},\ \bibinfo {pages} {2705} (\bibinfo {year} {1999})}\BibitemShut {NoStop}%
\bibitem [{\citenamefont {Savin}\ \emph {et~al.}(2001)\citenamefont {Savin}, \citenamefont {Prunnila}, \citenamefont {Kivinen}, \citenamefont {Pekola}, \citenamefont {Ahopelto},\ and\ \citenamefont {Manninen}}]{Savin2001}%
  \BibitemOpen
  \bibfield  {author} {\bibinfo {author} {\bibfnamefont {A.~M.}\ \bibnamefont {Savin}}, \bibinfo {author} {\bibfnamefont {M.}~\bibnamefont {Prunnila}}, \bibinfo {author} {\bibfnamefont {P.~P.}\ \bibnamefont {Kivinen}}, \bibinfo {author} {\bibfnamefont {J.~P.}\ \bibnamefont {Pekola}}, \bibinfo {author} {\bibfnamefont {J.}~\bibnamefont {Ahopelto}},\ and\ \bibinfo {author} {\bibfnamefont {A.~J.}\ \bibnamefont {Manninen}},\ }\bibfield  {title} {\bibinfo {title} {Efficient electronic cooling in heavily doped silicon by quasiparticle tunneling},\ }\href {https://doi.org/10.1063/1.1399313} {\bibfield  {journal} {\bibinfo  {journal} {Appl. Phys. Lett.}\ }\textbf {\bibinfo {volume} {79}},\ \bibinfo {pages} {1471} (\bibinfo {year} {2001})}\BibitemShut {NoStop}%
\bibitem [{\citenamefont {Pekola}\ \emph {et~al.}(2004)\citenamefont {Pekola}, \citenamefont {Heikkil\"a}, \citenamefont {Savin}, \citenamefont {Flyktman}, \citenamefont {Giazotto},\ and\ \citenamefont {Hekking}}]{Pekola2004}%
  \BibitemOpen
  \bibfield  {author} {\bibinfo {author} {\bibfnamefont {J.~P.}\ \bibnamefont {Pekola}}, \bibinfo {author} {\bibfnamefont {T.~T.}\ \bibnamefont {Heikkil\"a}}, \bibinfo {author} {\bibfnamefont {A.~M.}\ \bibnamefont {Savin}}, \bibinfo {author} {\bibfnamefont {J.~T.}\ \bibnamefont {Flyktman}}, \bibinfo {author} {\bibfnamefont {F.}~\bibnamefont {Giazotto}},\ and\ \bibinfo {author} {\bibfnamefont {F.~W.~J.}\ \bibnamefont {Hekking}},\ }\bibfield  {title} {\bibinfo {title} {Limitations in cooling electrons using normal-metal-superconductor tunnel junctions},\ }\href {https://doi.org/10.1103/PhysRevLett.92.056804} {\bibfield  {journal} {\bibinfo  {journal} {Phys. Rev. Lett.}\ }\textbf {\bibinfo {volume} {92}},\ \bibinfo {pages} {056804} (\bibinfo {year} {2004})}\BibitemShut {NoStop}%
\bibitem [{\citenamefont {Clark}\ \emph {et~al.}(2005)\citenamefont {Clark}, \citenamefont {Miller}, \citenamefont {Williams}, \citenamefont {Ruggiero}, \citenamefont {Hilton}, \citenamefont {Vale}, \citenamefont {Beall}, \citenamefont {Irwin},\ and\ \citenamefont {Ullom}}]{Clark2005}%
  \BibitemOpen
  \bibfield  {author} {\bibinfo {author} {\bibfnamefont {A.~M.}\ \bibnamefont {Clark}}, \bibinfo {author} {\bibfnamefont {N.~A.}\ \bibnamefont {Miller}}, \bibinfo {author} {\bibfnamefont {A.}~\bibnamefont {Williams}}, \bibinfo {author} {\bibfnamefont {S.~T.}\ \bibnamefont {Ruggiero}}, \bibinfo {author} {\bibfnamefont {G.~C.}\ \bibnamefont {Hilton}}, \bibinfo {author} {\bibfnamefont {L.~R.}\ \bibnamefont {Vale}}, \bibinfo {author} {\bibfnamefont {J.~A.}\ \bibnamefont {Beall}}, \bibinfo {author} {\bibfnamefont {K.~D.}\ \bibnamefont {Irwin}},\ and\ \bibinfo {author} {\bibfnamefont {J.~N.}\ \bibnamefont {Ullom}},\ }\bibfield  {title} {\bibinfo {title} {Cooling of bulk material by electron-tunneling refrigerators},\ }\href {https://doi.org/10.1063/1.1914966} {\bibfield  {journal} {\bibinfo  {journal} {Appl. Phys. Lett.}\ }\textbf {\bibinfo {volume} {86}},\ \bibinfo {pages} {173508} (\bibinfo {year} {2005})}\BibitemShut {NoStop}%
\bibitem [{\citenamefont {Vasenko}\ \emph {et~al.}(2010)\citenamefont {Vasenko}, \citenamefont {Bezuglyi}, \citenamefont {Courtois},\ and\ \citenamefont {Hekking}}]{Vasenko2010}%
  \BibitemOpen
  \bibfield  {author} {\bibinfo {author} {\bibfnamefont {A.~S.}\ \bibnamefont {Vasenko}}, \bibinfo {author} {\bibfnamefont {E.~V.}\ \bibnamefont {Bezuglyi}}, \bibinfo {author} {\bibfnamefont {H.}~\bibnamefont {Courtois}},\ and\ \bibinfo {author} {\bibfnamefont {F.~W.~J.}\ \bibnamefont {Hekking}},\ }\bibfield  {title} {\bibinfo {title} {Electron cooling by diffusive normal metal--superconductor tunnel junctions},\ }\href {https://doi.org/10.1103/PhysRevB.81.094513} {\bibfield  {journal} {\bibinfo  {journal} {Phys. Rev. B}\ }\textbf {\bibinfo {volume} {81}},\ \bibinfo {pages} {094513} (\bibinfo {year} {2010})}\BibitemShut {NoStop}%
\bibitem [{\citenamefont {Quaranta}\ \emph {et~al.}(2011)\citenamefont {Quaranta}, \citenamefont {Spathis}, \citenamefont {Beltram},\ and\ \citenamefont {Giazotto}}]{Quaranta2011}%
  \BibitemOpen
  \bibfield  {author} {\bibinfo {author} {\bibfnamefont {O.}~\bibnamefont {Quaranta}}, \bibinfo {author} {\bibfnamefont {P.}~\bibnamefont {Spathis}}, \bibinfo {author} {\bibfnamefont {F.}~\bibnamefont {Beltram}},\ and\ \bibinfo {author} {\bibfnamefont {F.}~\bibnamefont {Giazotto}},\ }\bibfield  {title} {\bibinfo {title} {{Cooling electrons from 1 to 0.4 K with V-based nanorefrigerators}},\ }\href {https://doi.org/10.1063/1.3544058} {\bibfield  {journal} {\bibinfo  {journal} {Appl. Phys. Lett.}\ }\textbf {\bibinfo {volume} {98}},\ \bibinfo {pages} {032501} (\bibinfo {year} {2011})}\BibitemShut {NoStop}%
\bibitem [{\citenamefont {Rajauria}\ \emph {et~al.}(2012)\citenamefont {Rajauria}, \citenamefont {Pascal}, \citenamefont {Gandit}, \citenamefont {Hekking}, \citenamefont {Pannetier},\ and\ \citenamefont {Courtois}}]{Rajauria2012}%
  \BibitemOpen
  \bibfield  {author} {\bibinfo {author} {\bibfnamefont {S.}~\bibnamefont {Rajauria}}, \bibinfo {author} {\bibfnamefont {L.~M.~A.}\ \bibnamefont {Pascal}}, \bibinfo {author} {\bibfnamefont {P.}~\bibnamefont {Gandit}}, \bibinfo {author} {\bibfnamefont {F.~W.~J.}\ \bibnamefont {Hekking}}, \bibinfo {author} {\bibfnamefont {B.}~\bibnamefont {Pannetier}},\ and\ \bibinfo {author} {\bibfnamefont {H.}~\bibnamefont {Courtois}},\ }\bibfield  {title} {\bibinfo {title} {Efficiency of quasiparticle evacuation in superconducting devices},\ }\href {https://doi.org/10.1103/PhysRevB.85.020505} {\bibfield  {journal} {\bibinfo  {journal} {Phys. Rev. B}\ }\textbf {\bibinfo {volume} {85}},\ \bibinfo {pages} {020505} (\bibinfo {year} {2012})}\BibitemShut {NoStop}%
\bibitem [{\citenamefont {Lowell}\ \emph {et~al.}(2013)\citenamefont {Lowell}, \citenamefont {O'Neil}, \citenamefont {Underwood},\ and\ \citenamefont {Ullom}}]{Lowell2013}%
  \BibitemOpen
  \bibfield  {author} {\bibinfo {author} {\bibfnamefont {P.~J.}\ \bibnamefont {Lowell}}, \bibinfo {author} {\bibfnamefont {G.~C.}\ \bibnamefont {O'Neil}}, \bibinfo {author} {\bibfnamefont {J.~M.}\ \bibnamefont {Underwood}},\ and\ \bibinfo {author} {\bibfnamefont {J.~N.}\ \bibnamefont {Ullom}},\ }\bibfield  {title} {\bibinfo {title} {Macroscale refrigeration by nanoscale electron transport},\ }\href {https://doi.org/10.1063/1.4793515} {\bibfield  {journal} {\bibinfo  {journal} {Appl. Phys. Lett.}\ }\textbf {\bibinfo {volume} {102}},\ \bibinfo {pages} {082601} (\bibinfo {year} {2013})}\BibitemShut {NoStop}%
\bibitem [{\citenamefont {Nguyen}\ \emph {et~al.}(2014)\citenamefont {Nguyen}, \citenamefont {Meschke}, \citenamefont {Courtois},\ and\ \citenamefont {Pekola}}]{Nguyen2014}%
  \BibitemOpen
  \bibfield  {author} {\bibinfo {author} {\bibfnamefont {H.~Q.}\ \bibnamefont {Nguyen}}, \bibinfo {author} {\bibfnamefont {M.}~\bibnamefont {Meschke}}, \bibinfo {author} {\bibfnamefont {H.}~\bibnamefont {Courtois}},\ and\ \bibinfo {author} {\bibfnamefont {J.~P.}\ \bibnamefont {Pekola}},\ }\bibfield  {title} {\bibinfo {title} {{Sub-50-mK electronic cooling with large-area superconducting tunnel junctions}},\ }\href {https://doi.org/10.1103/PhysRevApplied.2.054001} {\bibfield  {journal} {\bibinfo  {journal} {Phys. Rev. Appl.}\ }\textbf {\bibinfo {volume} {2}},\ \bibinfo {pages} {054001} (\bibinfo {year} {2014})}\BibitemShut {NoStop}%
\bibitem [{\citenamefont {Camarasa-G\'omez}\ \emph {et~al.}(2014)\citenamefont {Camarasa-G\'omez}, \citenamefont {Di~Marco}, \citenamefont {Hekking}, \citenamefont {Winkelmann}, \citenamefont {Courtois},\ and\ \citenamefont {Giazotto}}]{Camarasa2014}%
  \BibitemOpen
  \bibfield  {author} {\bibinfo {author} {\bibfnamefont {M.}~\bibnamefont {Camarasa-G\'omez}}, \bibinfo {author} {\bibfnamefont {A.}~\bibnamefont {Di~Marco}}, \bibinfo {author} {\bibfnamefont {F.~W.~J.}\ \bibnamefont {Hekking}}, \bibinfo {author} {\bibfnamefont {C.~B.}\ \bibnamefont {Winkelmann}}, \bibinfo {author} {\bibfnamefont {H.}~\bibnamefont {Courtois}},\ and\ \bibinfo {author} {\bibfnamefont {F.}~\bibnamefont {Giazotto}},\ }\bibfield  {title} {\bibinfo {title} {Superconducting cascade electron refrigerator},\ }\href {https://doi.org/10.1063/1.4876478} {\bibfield  {journal} {\bibinfo  {journal} {Appl. Phys. Lett.}\ }\textbf {\bibinfo {volume} {104}},\ \bibinfo {pages} {192601} (\bibinfo {year} {2014})}\BibitemShut {NoStop}%
\bibitem [{\citenamefont {Nguyen}\ \emph {et~al.}(2015)\citenamefont {Nguyen}, \citenamefont {Meschke},\ and\ \citenamefont {Pekola}}]{Nguyen2015}%
  \BibitemOpen
  \bibfield  {author} {\bibinfo {author} {\bibfnamefont {H.~Q.}\ \bibnamefont {Nguyen}}, \bibinfo {author} {\bibfnamefont {M.}~\bibnamefont {Meschke}},\ and\ \bibinfo {author} {\bibfnamefont {J.~P.}\ \bibnamefont {Pekola}},\ }\bibfield  {title} {\bibinfo {title} {A robust platform cooled by superconducting electronic refrigerators},\ }\href {https://doi.org/10.1063/1.4905440} {\bibfield  {journal} {\bibinfo  {journal} {Appl. Phys. Lett.}\ }\textbf {\bibinfo {volume} {106}},\ \bibinfo {pages} {012601} (\bibinfo {year} {2015})}\BibitemShut {NoStop}%
\bibitem [{\citenamefont {Zhang}\ \emph {et~al.}(2015)\citenamefont {Zhang}, \citenamefont {Lowell}, \citenamefont {Wilson}, \citenamefont {O'Neil},\ and\ \citenamefont {Ullom}}]{Zhang2015}%
  \BibitemOpen
  \bibfield  {author} {\bibinfo {author} {\bibfnamefont {X.}~\bibnamefont {Zhang}}, \bibinfo {author} {\bibfnamefont {P.~J.}\ \bibnamefont {Lowell}}, \bibinfo {author} {\bibfnamefont {B.~L.}\ \bibnamefont {Wilson}}, \bibinfo {author} {\bibfnamefont {G.~C.}\ \bibnamefont {O'Neil}},\ and\ \bibinfo {author} {\bibfnamefont {J.~N.}\ \bibnamefont {Ullom}},\ }\bibfield  {title} {\bibinfo {title} {Macroscopic subkelvin refrigerator employing superconducting tunnel junctions},\ }\href {https://doi.org/10.1103/PhysRevApplied.4.024006} {\bibfield  {journal} {\bibinfo  {journal} {Phys. Rev. Appl.}\ }\textbf {\bibinfo {volume} {4}},\ \bibinfo {pages} {024006} (\bibinfo {year} {2015})}\BibitemShut {NoStop}%
\bibitem [{\citenamefont {Gunnarsson}\ \emph {et~al.}(2015)\citenamefont {Gunnarsson}, \citenamefont {Richardson-Bullock}, \citenamefont {Prest}, \citenamefont {Nguyena}, \citenamefont {Timofeev}, \citenamefont {Shah}, \citenamefont {Whall}, \citenamefont {Parker}, \citenamefont {Leadley}, \citenamefont {Myronov},\ and\ \citenamefont {Prunnila}}]{Gunnarsson2015}%
  \BibitemOpen
  \bibfield  {author} {\bibinfo {author} {\bibfnamefont {D.}~\bibnamefont {Gunnarsson}}, \bibinfo {author} {\bibfnamefont {J.}~\bibnamefont {Richardson-Bullock}}, \bibinfo {author} {\bibfnamefont {M.}~\bibnamefont {Prest}}, \bibinfo {author} {\bibfnamefont {H.~Q.}\ \bibnamefont {Nguyena}}, \bibinfo {author} {\bibfnamefont {A.~V.}\ \bibnamefont {Timofeev}}, \bibinfo {author} {\bibfnamefont {V.~A.}\ \bibnamefont {Shah}}, \bibinfo {author} {\bibfnamefont {T.~E.}\ \bibnamefont {Whall}}, \bibinfo {author} {\bibfnamefont {E.~H.~C.}\ \bibnamefont {Parker}}, \bibinfo {author} {\bibfnamefont {D.~R.}\ \bibnamefont {Leadley}}, \bibinfo {author} {\bibfnamefont {M.}~\bibnamefont {Myronov}},\ and\ \bibinfo {author} {\bibfnamefont {M.}~\bibnamefont {Prunnila}},\ }\bibfield  {title} {\bibinfo {title} {Interfacial engineering of semiconductor–superconductor junctions for high performance micro-coolers},\ }\href {https://doi.org/10.1038/srep17398} {\bibfield  {journal} {\bibinfo  {journal} {Sci. Rep.}\ }\textbf {\bibinfo
  {volume} {5}},\ \bibinfo {pages} {17398} (\bibinfo {year} {2015})}\BibitemShut {NoStop}%
\bibitem [{\citenamefont {Nguyen}\ \emph {et~al.}(2016)\citenamefont {Nguyen}, \citenamefont {Peltonen}, \citenamefont {Meschke},\ and\ \citenamefont {Pekola}}]{Nguyen2016}%
  \BibitemOpen
  \bibfield  {author} {\bibinfo {author} {\bibfnamefont {H.~Q.}\ \bibnamefont {Nguyen}}, \bibinfo {author} {\bibfnamefont {J.~T.}\ \bibnamefont {Peltonen}}, \bibinfo {author} {\bibfnamefont {M.}~\bibnamefont {Meschke}},\ and\ \bibinfo {author} {\bibfnamefont {J.~P.}\ \bibnamefont {Pekola}},\ }\bibfield  {title} {\bibinfo {title} {Cascade electronic refrigerator using superconducting tunnel junctions},\ }\href {https://doi.org/10.1103/PhysRevApplied.6.054011} {\bibfield  {journal} {\bibinfo  {journal} {Phys. Rev. Appl.}\ }\textbf {\bibinfo {volume} {6}},\ \bibinfo {pages} {054011} (\bibinfo {year} {2016})}\BibitemShut {NoStop}%
\bibitem [{\citenamefont {Kashiwaya}\ \emph {et~al.}(2016)\citenamefont {Kashiwaya}, \citenamefont {Kashiwaya}, \citenamefont {Koyanagi},\ and\ \citenamefont {Tanaka}}]{Kashiwaya2016}%
  \BibitemOpen
  \bibfield  {author} {\bibinfo {author} {\bibfnamefont {S.}~\bibnamefont {Kashiwaya}}, \bibinfo {author} {\bibfnamefont {H.}~\bibnamefont {Kashiwaya}}, \bibinfo {author} {\bibfnamefont {M.}~\bibnamefont {Koyanagi}},\ and\ \bibinfo {author} {\bibfnamefont {Y.}~\bibnamefont {Tanaka}},\ }\bibfield  {title} {\bibinfo {title} {Development of suspended normal-metal-type tunneling junction refrigerator},\ }\href {https://doi.org/10.7567/JJAP.55.093101} {\bibfield  {journal} {\bibinfo  {journal} {Jpn. J. Appl. Phys.}\ }\textbf {\bibinfo {volume} {55}},\ \bibinfo {pages} {093101} (\bibinfo {year} {2016})}\BibitemShut {NoStop}%
\bibitem [{\citenamefont {Courtois}\ \emph {et~al.}(2016)\citenamefont {Courtois}, \citenamefont {Nguyen}, \citenamefont {Winkelmann},\ and\ \citenamefont {Pekola}}]{Courtois2016}%
  \BibitemOpen
  \bibfield  {author} {\bibinfo {author} {\bibfnamefont {H.}~\bibnamefont {Courtois}}, \bibinfo {author} {\bibfnamefont {H.~Q.}\ \bibnamefont {Nguyen}}, \bibinfo {author} {\bibfnamefont {C.~B.}\ \bibnamefont {Winkelmann}},\ and\ \bibinfo {author} {\bibfnamefont {J.~P.}\ \bibnamefont {Pekola}},\ }\bibfield  {title} {\bibinfo {title} {{High-performance electronic cooling with superconducting tunnel junctions}},\ }\href {https://doi.org/10.1016/j.crhy.2016.08.010} {\bibfield  {journal} {\bibinfo  {journal} {Comptes Rendus. Physique}\ }\textbf {\bibinfo {volume} {17}},\ \bibinfo {pages} {1139} (\bibinfo {year} {2016})}\BibitemShut {NoStop}%
\bibitem [{\citenamefont {S\'anchez}(2017)}]{Sanchez2017}%
  \BibitemOpen
  \bibfield  {author} {\bibinfo {author} {\bibfnamefont {R.}~\bibnamefont {S\'anchez}},\ }\bibfield  {title} {\bibinfo {title} {Correlation-induced refrigeration with superconducting single-electron transistors},\ }\href {https://doi.org/10.1063/1.5008481} {\bibfield  {journal} {\bibinfo  {journal} {Appl. Phys. Lett.}\ }\textbf {\bibinfo {volume} {111}},\ \bibinfo {pages} {223103} (\bibinfo {year} {2017})}\BibitemShut {NoStop}%
\bibitem [{\citenamefont {S\'anchez}\ \emph {et~al.}(2018)\citenamefont {S\'anchez}, \citenamefont {Burset},\ and\ \citenamefont {Yeyati}}]{Sanchez2018}%
  \BibitemOpen
  \bibfield  {author} {\bibinfo {author} {\bibfnamefont {R.}~\bibnamefont {S\'anchez}}, \bibinfo {author} {\bibfnamefont {P.}~\bibnamefont {Burset}},\ and\ \bibinfo {author} {\bibfnamefont {A.~L.}\ \bibnamefont {Yeyati}},\ }\bibfield  {title} {\bibinfo {title} {{Cooling by Cooper pair splitting}},\ }\href {https://doi.org/10.1103/PhysRevB.98.241414} {\bibfield  {journal} {\bibinfo  {journal} {Phys. Rev. B}\ }\textbf {\bibinfo {volume} {98}},\ \bibinfo {pages} {241414} (\bibinfo {year} {2018})}\BibitemShut {NoStop}%
\bibitem [{\citenamefont {Marchegiani}\ \emph {et~al.}(2018)\citenamefont {Marchegiani}, \citenamefont {Virtanen},\ and\ \citenamefont {Giazotto}}]{Marchegiani2018}%
  \BibitemOpen
  \bibfield  {author} {\bibinfo {author} {\bibfnamefont {G.}~\bibnamefont {Marchegiani}}, \bibinfo {author} {\bibfnamefont {P.}~\bibnamefont {Virtanen}},\ and\ \bibinfo {author} {\bibfnamefont {F.}~\bibnamefont {Giazotto}},\ }\bibfield  {title} {\bibinfo {title} {On-chip cooling by heating with superconducting tunnel junctions},\ }\href {https://doi.org/10.1209/0295-5075/124/48005} {\bibfield  {journal} {\bibinfo  {journal} {Europhysics Letters}\ }\textbf {\bibinfo {volume} {124}},\ \bibinfo {pages} {48005} (\bibinfo {year} {2018})}\BibitemShut {NoStop}%
\bibitem [{\citenamefont {Kapri}\ and\ \citenamefont {Basu}(2018)}]{Kapri2018}%
  \BibitemOpen
  \bibfield  {author} {\bibinfo {author} {\bibfnamefont {P.}~\bibnamefont {Kapri}}\ and\ \bibinfo {author} {\bibfnamefont {S.}~\bibnamefont {Basu}},\ }\bibfield  {title} {\bibinfo {title} {Tunable refrigeration properties of nano-scale rashba coupled junction devices},\ }\href {https://doi.org/https://doi.org/10.1016/j.physe.2018.06.018} {\bibfield  {journal} {\bibinfo  {journal} {Physica E: Low-dimensional Systems and Nanostructures}\ }\textbf {\bibinfo {volume} {103}},\ \bibinfo {pages} {383} (\bibinfo {year} {2018})}\BibitemShut {NoStop}%
\bibitem [{\citenamefont {Kapri}\ and\ \citenamefont {Basu}(2019)}]{Kapri2019}%
  \BibitemOpen
  \bibfield  {author} {\bibinfo {author} {\bibfnamefont {P.}~\bibnamefont {Kapri}}\ and\ \bibinfo {author} {\bibfnamefont {S.}~\bibnamefont {Basu}},\ }\bibfield  {title} {\bibinfo {title} {Thermopower generation and thermoelectric cooling in a kane-mele normal-insulator-superconductor nano-junction},\ }\href {https://doi.org/10.1209/0295-5075/125/47003} {\bibfield  {journal} {\bibinfo  {journal} {Europhysics Letters}\ }\textbf {\bibinfo {volume} {125}},\ \bibinfo {pages} {47003} (\bibinfo {year} {2019})}\BibitemShut {NoStop}%
\bibitem [{\citenamefont {Hussein}\ \emph {et~al.}(2019)\citenamefont {Hussein}, \citenamefont {Governale}, \citenamefont {Kohler}, \citenamefont {Belzig}, \citenamefont {Giazotto},\ and\ \citenamefont {Braggio}}]{Hussein2019}%
  \BibitemOpen
  \bibfield  {author} {\bibinfo {author} {\bibfnamefont {R.}~\bibnamefont {Hussein}}, \bibinfo {author} {\bibfnamefont {M.}~\bibnamefont {Governale}}, \bibinfo {author} {\bibfnamefont {S.}~\bibnamefont {Kohler}}, \bibinfo {author} {\bibfnamefont {W.}~\bibnamefont {Belzig}}, \bibinfo {author} {\bibfnamefont {F.}~\bibnamefont {Giazotto}},\ and\ \bibinfo {author} {\bibfnamefont {A.}~\bibnamefont {Braggio}},\ }\bibfield  {title} {\bibinfo {title} {{Nonlocal thermoelectricity in a Cooper-pair splitter}},\ }\href {https://doi.org/10.1103/PhysRevB.99.075429} {\bibfield  {journal} {\bibinfo  {journal} {Phys. Rev. B}\ }\textbf {\bibinfo {volume} {99}},\ \bibinfo {pages} {075429} (\bibinfo {year} {2019})}\BibitemShut {NoStop}%
\bibitem [{\citenamefont {Gordeeva}\ \emph {et~al.}(2020)\citenamefont {Gordeeva}, \citenamefont {Pankratov}, \citenamefont {Pugach}, \citenamefont {Vasenko}, \citenamefont {Zbrozhek}, \citenamefont {Blagodatkin}, \citenamefont {Pimanov},\ and\ \citenamefont {Kuzmin}}]{Gordeeva2020}%
  \BibitemOpen
  \bibfield  {author} {\bibinfo {author} {\bibfnamefont {A.~V.}\ \bibnamefont {Gordeeva}}, \bibinfo {author} {\bibfnamefont {A.~L.}\ \bibnamefont {Pankratov}}, \bibinfo {author} {\bibfnamefont {N.~G.}\ \bibnamefont {Pugach}}, \bibinfo {author} {\bibfnamefont {A.~S.}\ \bibnamefont {Vasenko}}, \bibinfo {author} {\bibfnamefont {V.~O.}\ \bibnamefont {Zbrozhek}}, \bibinfo {author} {\bibfnamefont {A.~V.}\ \bibnamefont {Blagodatkin}}, \bibinfo {author} {\bibfnamefont {D.~A.}\ \bibnamefont {Pimanov}},\ and\ \bibinfo {author} {\bibfnamefont {L.~S.}\ \bibnamefont {Kuzmin}},\ }\bibfield  {title} {\bibinfo {title} {Record electron self-cooling in cold-electron bolometers with a hybrid superconductor-ferromagnetic nanoabsorber and traps},\ }\href {https://doi.org/10.1038/s41598-020-78869-z} {\bibfield  {journal} {\bibinfo  {journal} {Sci. Rep.}\ }\textbf {\bibinfo {volume} {10}},\ \bibinfo {pages} {21961} (\bibinfo {year} {2020})}\BibitemShut {NoStop}%
\bibitem [{\citenamefont {Vischi}\ \emph {et~al.}(2020)\citenamefont {Vischi}, \citenamefont {Carrega}, \citenamefont {Braggio}, \citenamefont {Paolucci}, \citenamefont {Bianco}, \citenamefont {Roddaro},\ and\ \citenamefont {Giazotto}}]{Vischi2020}%
  \BibitemOpen
  \bibfield  {author} {\bibinfo {author} {\bibfnamefont {F.}~\bibnamefont {Vischi}}, \bibinfo {author} {\bibfnamefont {M.}~\bibnamefont {Carrega}}, \bibinfo {author} {\bibfnamefont {A.}~\bibnamefont {Braggio}}, \bibinfo {author} {\bibfnamefont {F.}~\bibnamefont {Paolucci}}, \bibinfo {author} {\bibfnamefont {F.}~\bibnamefont {Bianco}}, \bibinfo {author} {\bibfnamefont {S.}~\bibnamefont {Roddaro}},\ and\ \bibinfo {author} {\bibfnamefont {F.}~\bibnamefont {Giazotto}},\ }\bibfield  {title} {\bibinfo {title} {Electron cooling with graphene-insulator-superconductor tunnel junctions for applications in fast bolometry},\ }\href {https://doi.org/10.1103/PhysRevApplied.13.054006} {\bibfield  {journal} {\bibinfo  {journal} {Phys. Rev. Appl.}\ }\textbf {\bibinfo {volume} {13}},\ \bibinfo {pages} {054006} (\bibinfo {year} {2020})}\BibitemShut {NoStop}%
\bibitem [{\citenamefont {Kemppinen}\ \emph {et~al.}(2021)\citenamefont {Kemppinen}, \citenamefont {Ronzani}, \citenamefont {Mykkänen}, \citenamefont {Hätinen}, \citenamefont {Lehtinen},\ and\ \citenamefont {Prunnila}}]{Kemppinen2021}%
  \BibitemOpen
  \bibfield  {author} {\bibinfo {author} {\bibfnamefont {A.}~\bibnamefont {Kemppinen}}, \bibinfo {author} {\bibfnamefont {A.}~\bibnamefont {Ronzani}}, \bibinfo {author} {\bibfnamefont {E.}~\bibnamefont {Mykkänen}}, \bibinfo {author} {\bibfnamefont {J.}~\bibnamefont {Hätinen}}, \bibinfo {author} {\bibfnamefont {J.~S.}\ \bibnamefont {Lehtinen}},\ and\ \bibinfo {author} {\bibfnamefont {M.}~\bibnamefont {Prunnila}},\ }\bibfield  {title} {\bibinfo {title} {Cascaded superconducting junction refrigerators: Optimization and performance limits},\ }\href {https://doi.org/10.1063/5.0060652} {\bibfield  {journal} {\bibinfo  {journal} {Appl. Phys. Lett.}\ }\textbf {\bibinfo {volume} {119}},\ \bibinfo {pages} {052603} (\bibinfo {year} {2021})}\BibitemShut {NoStop}%
\bibitem [{\citenamefont {Hwang}\ \emph {et~al.}(2023)\citenamefont {Hwang}, \citenamefont {Sothmann},\ and\ \citenamefont {S\'anchez}}]{Hwang2023}%
  \BibitemOpen
  \bibfield  {author} {\bibinfo {author} {\bibfnamefont {S.-Y.}\ \bibnamefont {Hwang}}, \bibinfo {author} {\bibfnamefont {B.}~\bibnamefont {Sothmann}},\ and\ \bibinfo {author} {\bibfnamefont {D.}~\bibnamefont {S\'anchez}},\ }\bibfield  {title} {\bibinfo {title} {Superconductor--quantum dot hybrid coolers},\ }\href {https://doi.org/10.1103/PhysRevB.107.245412} {\bibfield  {journal} {\bibinfo  {journal} {Phys. Rev. B}\ }\textbf {\bibinfo {volume} {107}},\ \bibinfo {pages} {245412} (\bibinfo {year} {2023})}\BibitemShut {NoStop}%
\bibitem [{\citenamefont {H\"atinen}\ \emph {et~al.}(2024)\citenamefont {H\"atinen}, \citenamefont {Ronzani}, \citenamefont {Loreto}, \citenamefont {Mykk\"anen}, \citenamefont {Kemppinen}, \citenamefont {Viisanen}, \citenamefont {Rantanen}, \citenamefont {Geisor}, \citenamefont {Lehtinen}, \citenamefont {Ribeiro}, \citenamefont {Kaikkonen}, \citenamefont {Prakash}, \citenamefont {Vesterinen}, \citenamefont {F\"orbom}, \citenamefont {Mannila}, \citenamefont {Kervinen}, \citenamefont {Govenius},\ and\ \citenamefont {Prunnila}}]{Hatinen2024}%
  \BibitemOpen
  \bibfield  {author} {\bibinfo {author} {\bibfnamefont {J.}~\bibnamefont {H\"atinen}}, \bibinfo {author} {\bibfnamefont {A.}~\bibnamefont {Ronzani}}, \bibinfo {author} {\bibfnamefont {R.}~\bibnamefont {Loreto}}, \bibinfo {author} {\bibfnamefont {E.}~\bibnamefont {Mykk\"anen}}, \bibinfo {author} {\bibfnamefont {A.}~\bibnamefont {Kemppinen}}, \bibinfo {author} {\bibfnamefont {K.}~\bibnamefont {Viisanen}}, \bibinfo {author} {\bibfnamefont {T.}~\bibnamefont {Rantanen}}, \bibinfo {author} {\bibfnamefont {J.}~\bibnamefont {Geisor}}, \bibinfo {author} {\bibfnamefont {J.}~\bibnamefont {Lehtinen}}, \bibinfo {author} {\bibfnamefont {M.}~\bibnamefont {Ribeiro}}, \bibinfo {author} {\bibfnamefont {J.-P.}\ \bibnamefont {Kaikkonen}}, \bibinfo {author} {\bibfnamefont {O.}~\bibnamefont {Prakash}}, \bibinfo {author} {\bibfnamefont {V.}~\bibnamefont {Vesterinen}}, \bibinfo {author} {\bibfnamefont {C.}~\bibnamefont {F\"orbom}}, \bibinfo {author} {\bibfnamefont {E.}~\bibnamefont {Mannila}}, \bibinfo {author} {\bibfnamefont
  {M.}~\bibnamefont {Kervinen}}, \bibinfo {author} {\bibfnamefont {J.}~\bibnamefont {Govenius}},\ and\ \bibinfo {author} {\bibfnamefont {M.}~\bibnamefont {Prunnila}},\ }\bibfield  {title} {\bibinfo {title} {Efficient electronic cooling by niobium-based superconducting tunnel junctions},\ }\href {https://doi.org/10.1103/PhysRevApplied.22.064048} {\bibfield  {journal} {\bibinfo  {journal} {Phys. Rev. Appl.}\ }\textbf {\bibinfo {volume} {22}},\ \bibinfo {pages} {064048} (\bibinfo {year} {2024})}\BibitemShut {NoStop}%
\bibitem [{\citenamefont {Verma}\ and\ \citenamefont {Singh}(2024)}]{Verma2024}%
  \BibitemOpen
  \bibfield  {author} {\bibinfo {author} {\bibfnamefont {S.}~\bibnamefont {Verma}}\ and\ \bibinfo {author} {\bibfnamefont {A.}~\bibnamefont {Singh}},\ }\bibfield  {title} {\bibinfo {title} {Seebeck power generation and peltier cooling in a normal metal-quantum dot-superconductor nanodevice},\ }\href {https://doi.org/10.1007/s10909-024-03047-8} {\bibfield  {journal} {\bibinfo  {journal} {J. Low Temp. Phys.}\ }\textbf {\bibinfo {volume} {214}},\ \bibinfo {pages} {344} (\bibinfo {year} {2024})}\BibitemShut {NoStop}%
\bibitem [{\citenamefont {Giazotto}\ \emph {et~al.}(2002)\citenamefont {Giazotto}, \citenamefont {Taddei}, \citenamefont {Fazio},\ and\ \citenamefont {Beltram}}]{Giazotto2002}%
  \BibitemOpen
  \bibfield  {author} {\bibinfo {author} {\bibfnamefont {F.}~\bibnamefont {Giazotto}}, \bibinfo {author} {\bibfnamefont {F.}~\bibnamefont {Taddei}}, \bibinfo {author} {\bibfnamefont {R.}~\bibnamefont {Fazio}},\ and\ \bibinfo {author} {\bibfnamefont {F.~M.}\ \bibnamefont {Beltram}},\ }\bibfield  {title} {\bibinfo {title} {Ultraefficient cooling in ferromagnet–superconductor microrefrigerators},\ }\href {https://doi.org/10.1063/1.1481242} {\bibfield  {journal} {\bibinfo  {journal} {App. Phys. Lett.}\ }\textbf {\bibinfo {volume} {80}},\ \bibinfo {pages} {3784} (\bibinfo {year} {2002})}\BibitemShut {NoStop}%
\bibitem [{\citenamefont {Giazotto}\ \emph {et~al.}(2006{\natexlab{b}})\citenamefont {Giazotto}, \citenamefont {Taddei}, \citenamefont {Governale}, \citenamefont {Castellana}, \citenamefont {Fazio},\ and\ \citenamefont {Beltram}}]{Giazotto2006b}%
  \BibitemOpen
  \bibfield  {author} {\bibinfo {author} {\bibfnamefont {F.}~\bibnamefont {Giazotto}}, \bibinfo {author} {\bibfnamefont {F.}~\bibnamefont {Taddei}}, \bibinfo {author} {\bibfnamefont {M.}~\bibnamefont {Governale}}, \bibinfo {author} {\bibfnamefont {C.}~\bibnamefont {Castellana}}, \bibinfo {author} {\bibfnamefont {R.}~\bibnamefont {Fazio}},\ and\ \bibinfo {author} {\bibfnamefont {F.}~\bibnamefont {Beltram}},\ }\bibfield  {title} {\bibinfo {title} {{Cooling electrons by magnetic-field tuning of Andreev reflection}},\ }\href {https://doi.org/10.1103/PhysRevLett.97.197001} {\bibfield  {journal} {\bibinfo  {journal} {Phys. Rev. Lett.}\ }\textbf {\bibinfo {volume} {97}},\ \bibinfo {pages} {197001} (\bibinfo {year} {2006}{\natexlab{b}})}\BibitemShut {NoStop}%
\bibitem [{\citenamefont {Ozaeta}\ \emph {et~al.}(2012)\citenamefont {Ozaeta}, \citenamefont {Vasenko}, \citenamefont {Hekking},\ and\ \citenamefont {Bergeret}}]{Ozaeta2012}%
  \BibitemOpen
  \bibfield  {author} {\bibinfo {author} {\bibfnamefont {A.}~\bibnamefont {Ozaeta}}, \bibinfo {author} {\bibfnamefont {A.~S.}\ \bibnamefont {Vasenko}}, \bibinfo {author} {\bibfnamefont {F.~W.~J.}\ \bibnamefont {Hekking}},\ and\ \bibinfo {author} {\bibfnamefont {F.~S.}\ \bibnamefont {Bergeret}},\ }\bibfield  {title} {\bibinfo {title} {Electron cooling in diffusive normal metal--superconductor tunnel junctions with a spin-valve ferromagnetic interlayer},\ }\href {https://doi.org/10.1103/PhysRevB.85.174518} {\bibfield  {journal} {\bibinfo  {journal} {Phys. Rev. B}\ }\textbf {\bibinfo {volume} {85}},\ \bibinfo {pages} {174518} (\bibinfo {year} {2012})}\BibitemShut {NoStop}%
\bibitem [{\citenamefont {Kawabata}\ \emph {et~al.}(2013)\citenamefont {Kawabata}, \citenamefont {Ozaeta}, \citenamefont {Vasenko}, \citenamefont {Hekking},\ and\ \citenamefont {Sebasti\'an~Bergeret}}]{Kawabata2013}%
  \BibitemOpen
  \bibfield  {author} {\bibinfo {author} {\bibfnamefont {S.}~\bibnamefont {Kawabata}}, \bibinfo {author} {\bibfnamefont {A.}~\bibnamefont {Ozaeta}}, \bibinfo {author} {\bibfnamefont {A.~S.}\ \bibnamefont {Vasenko}}, \bibinfo {author} {\bibfnamefont {F.~W.~J.}\ \bibnamefont {Hekking}},\ and\ \bibinfo {author} {\bibfnamefont {F.}~\bibnamefont {Sebasti\'an~Bergeret}},\ }\bibfield  {title} {\bibinfo {title} {Efficient electron refrigeration using superconductor/spin-filter devices},\ }\href {https://doi.org/10.1063/1.4813599} {\bibfield  {journal} {\bibinfo  {journal} {Appl. Phys. Lett.}\ }\textbf {\bibinfo {volume} {103}},\ \bibinfo {pages} {032602} (\bibinfo {year} {2013})}\BibitemShut {NoStop}%
\bibitem [{\citenamefont {Rouco}\ \emph {et~al.}(2018)\citenamefont {Rouco}, \citenamefont {Heikkil\"a},\ and\ \citenamefont {Bergeret}}]{Rouco2018}%
  \BibitemOpen
  \bibfield  {author} {\bibinfo {author} {\bibfnamefont {M.}~\bibnamefont {Rouco}}, \bibinfo {author} {\bibfnamefont {T.~T.}\ \bibnamefont {Heikkil\"a}},\ and\ \bibinfo {author} {\bibfnamefont {F.~S.}\ \bibnamefont {Bergeret}},\ }\bibfield  {title} {\bibinfo {title} {Electron refrigeration in hybrid structures with spin-split superconductors},\ }\href {https://doi.org/10.1103/PhysRevB.97.014529} {\bibfield  {journal} {\bibinfo  {journal} {Phys. Rev. B}\ }\textbf {\bibinfo {volume} {97}},\ \bibinfo {pages} {014529} (\bibinfo {year} {2018})}\BibitemShut {NoStop}%
\bibitem [{\citenamefont {Pendry}(1983)}]{Pendry1983}%
  \BibitemOpen
  \bibfield  {author} {\bibinfo {author} {\bibfnamefont {J.~B.}\ \bibnamefont {Pendry}},\ }\bibfield  {title} {\bibinfo {title} {Quantum limits to the flow of information and entropy},\ }\href {https://doi.org/10.1088/0305-4470/16/10/012} {\bibfield  {journal} {\bibinfo  {journal} {Journal of Physics A: Mathematical and General}\ }\textbf {\bibinfo {volume} {16}},\ \bibinfo {pages} {2161} (\bibinfo {year} {1983})}\BibitemShut {NoStop}%
\bibitem [{\citenamefont {Whitney}(2013)}]{Whitney2013}%
  \BibitemOpen
  \bibfield  {author} {\bibinfo {author} {\bibfnamefont {R.~S.}\ \bibnamefont {Whitney}},\ }\bibfield  {title} {\bibinfo {title} {{Thermodynamic and quantum bounds on nonlinear dc thermoelectric transport}},\ }\href {https://doi.org/https://doi.org/10.1103/PhysRevB.87.115404} {\bibfield  {journal} {\bibinfo  {journal} {Phys. Rev. B}\ }\textbf {\bibinfo {volume} {87}},\ \bibinfo {pages} {115404} (\bibinfo {year} {2013})}\BibitemShut {NoStop}%
\bibitem [{\citenamefont {Datta}(1997)}]{Datta1997}%
  \BibitemOpen
  \bibfield  {author} {\bibinfo {author} {\bibfnamefont {S.}~\bibnamefont {Datta}},\ }\href {https://books.google.it/books?id=28BC-ofEhvUC} {\emph {\bibinfo {title} {Electronic Transport in Mesoscopic Systems}}},\ Cambridge Studies in Semiconductor Physi\ (\bibinfo  {publisher} {Cambridge University Press},\ \bibinfo {year} {1997})\BibitemShut {NoStop}%
\bibitem [{\citenamefont {B\"uttiker}\ \emph {et~al.}(1984)\citenamefont {B\"uttiker}, \citenamefont {Imry},\ and\ \citenamefont {Azbel}}]{Buttiker1984}%
  \BibitemOpen
  \bibfield  {author} {\bibinfo {author} {\bibfnamefont {M.}~\bibnamefont {B\"uttiker}}, \bibinfo {author} {\bibfnamefont {Y.}~\bibnamefont {Imry}},\ and\ \bibinfo {author} {\bibfnamefont {M.~Y.}\ \bibnamefont {Azbel}},\ }\bibfield  {title} {\bibinfo {title} {Quantum oscillations in one-dimensional normal-metal rings},\ }\href {https://doi.org/10.1103/PhysRevA.30.1982} {\bibfield  {journal} {\bibinfo  {journal} {Phys. Rev. A}\ }\textbf {\bibinfo {volume} {30}},\ \bibinfo {pages} {1982} (\bibinfo {year} {1984})}\BibitemShut {NoStop}%
\bibitem [{\citenamefont {Tinkham}(2004)}]{Tinkham2004}%
  \BibitemOpen
  \bibfield  {author} {\bibinfo {author} {\bibfnamefont {M.}~\bibnamefont {Tinkham}},\ }\href@noop {} {\emph {\bibinfo {title} {Introduction to superconductivity}}},\ Vol.~\bibinfo {volume} {1}\ (\bibinfo  {publisher} {Courier Corporation},\ \bibinfo {year} {2004})\BibitemShut {NoStop}%
\bibitem [{\citenamefont {Roukes}\ \emph {et~al.}(1985)\citenamefont {Roukes}, \citenamefont {Freeman}, \citenamefont {Germain}, \citenamefont {Richardson},\ and\ \citenamefont {Ketchen}}]{Roukes1985}%
  \BibitemOpen
  \bibfield  {author} {\bibinfo {author} {\bibfnamefont {M.~L.}\ \bibnamefont {Roukes}}, \bibinfo {author} {\bibfnamefont {M.~R.}\ \bibnamefont {Freeman}}, \bibinfo {author} {\bibfnamefont {R.~S.}\ \bibnamefont {Germain}}, \bibinfo {author} {\bibfnamefont {R.~C.}\ \bibnamefont {Richardson}},\ and\ \bibinfo {author} {\bibfnamefont {M.~B.}\ \bibnamefont {Ketchen}},\ }\bibfield  {title} {\bibinfo {title} {Hot electrons and energy transport in metals at millikelvin temperatures},\ }\href {https://doi.org/10.1103/PhysRevLett.55.422} {\bibfield  {journal} {\bibinfo  {journal} {Phys. Rev. Lett.}\ }\textbf {\bibinfo {volume} {55}},\ \bibinfo {pages} {422} (\bibinfo {year} {1985})}\BibitemShut {NoStop}%
\bibitem [{\citenamefont {Wellstood}\ \emph {et~al.}(1994)\citenamefont {Wellstood}, \citenamefont {Urbina},\ and\ \citenamefont {Clarke}}]{Wellstood1994}%
  \BibitemOpen
  \bibfield  {author} {\bibinfo {author} {\bibfnamefont {F.~C.}\ \bibnamefont {Wellstood}}, \bibinfo {author} {\bibfnamefont {C.}~\bibnamefont {Urbina}},\ and\ \bibinfo {author} {\bibfnamefont {J.}~\bibnamefont {Clarke}},\ }\bibfield  {title} {\bibinfo {title} {Hot-electron effects in metals},\ }\href {https://doi.org/10.1103/PhysRevB.49.5942} {\bibfield  {journal} {\bibinfo  {journal} {Phys. Rev. B}\ }\textbf {\bibinfo {volume} {49}},\ \bibinfo {pages} {5942} (\bibinfo {year} {1994})}\BibitemShut {NoStop}%
\bibitem [{Note1()}]{Note1}%
  \BibitemOpen
  \bibinfo {note} {One can check that the curves in Fig.~\ref {JvsT}(c), once normalized to the normal state resistance of the junction $R_{\protect \rm t}$, exactly match the ones where the cooling power is calculated with the tunneling formula, see Ref.~\cite {Giazotto2006}.}\BibitemShut {Stop}%
\bibitem [{\citenamefont {Giazotto}\ \emph {et~al.}(2007)\citenamefont {Giazotto}, \citenamefont {Taddei}, \citenamefont {Governale}, \citenamefont {Fazio},\ and\ \citenamefont {Beltram}}]{Giazotto2007}%
  \BibitemOpen
  \bibfield  {author} {\bibinfo {author} {\bibfnamefont {F.}~\bibnamefont {Giazotto}}, \bibinfo {author} {\bibfnamefont {F.}~\bibnamefont {Taddei}}, \bibinfo {author} {\bibfnamefont {M.}~\bibnamefont {Governale}}, \bibinfo {author} {\bibfnamefont {R.}~\bibnamefont {Fazio}},\ and\ \bibinfo {author} {\bibfnamefont {F.}~\bibnamefont {Beltram}},\ }\bibfield  {title} {\bibinfo {title} {Landau cooling in metal-semiconductor nanostructures},\ }\href {https://doi.org/10.1088/1367-2630/9/12/439} {\bibfield  {journal} {\bibinfo  {journal} {New J. Phys.}\ }\textbf {\bibinfo {volume} {9}},\ \bibinfo {pages} {439} (\bibinfo {year} {2007})}\BibitemShut {NoStop}%
\bibitem [{\citenamefont {Whitney}(2014)}]{whitney2014most}%
  \BibitemOpen
  \bibfield  {author} {\bibinfo {author} {\bibfnamefont {R.~S.}\ \bibnamefont {Whitney}},\ }\bibfield  {title} {\bibinfo {title} {Most efficient quantum thermoelectric at finite power output},\ }\href {https://doi.org/https://doi.org/10.1103/PhysRevLett.112.130601} {\bibfield  {journal} {\bibinfo  {journal} {Phys. Rev. Lett.}\ }\textbf {\bibinfo {volume} {112}},\ \bibinfo {pages} {130601} (\bibinfo {year} {2014})}\BibitemShut {NoStop}%
\bibitem [{\citenamefont {Whitney}(2015)}]{whitney2015finding}%
  \BibitemOpen
  \bibfield  {author} {\bibinfo {author} {\bibfnamefont {R.~S.}\ \bibnamefont {Whitney}},\ }\bibfield  {title} {\bibinfo {title} {Finding the quantum thermoelectric with maximal efficiency and minimal entropy production at given power output},\ }\href {https://doi.org/https://doi.org/10.1103/PhysRevB.91.115425} {\bibfield  {journal} {\bibinfo  {journal} {Phys. Rev. B}\ }\textbf {\bibinfo {volume} {91}},\ \bibinfo {pages} {115425} (\bibinfo {year} {2015})}\BibitemShut {NoStop}%
\bibitem [{\citenamefont {Claughton}\ and\ \citenamefont {Lambert}(1996)}]{Claughton1996}%
  \BibitemOpen
  \bibfield  {author} {\bibinfo {author} {\bibfnamefont {N.~R.}\ \bibnamefont {Claughton}}\ and\ \bibinfo {author} {\bibfnamefont {C.~J.}\ \bibnamefont {Lambert}},\ }\bibfield  {title} {\bibinfo {title} {Thermoelectric properties of mesoscopic superconductors},\ }\href {https://doi.org/10.1103/PhysRevB.53.6605} {\bibfield  {journal} {\bibinfo  {journal} {Phys. Rev. B}\ }\textbf {\bibinfo {volume} {53}},\ \bibinfo {pages} {6605} (\bibinfo {year} {1996})}\BibitemShut {NoStop}%
\bibitem [{\citenamefont {M\'elin}\ \emph {et~al.}(2024)\citenamefont {M\'elin}, \citenamefont {Rashid},\ and\ \citenamefont {Kayyalha}}]{Melin2024}%
  \BibitemOpen
  \bibfield  {author} {\bibinfo {author} {\bibfnamefont {R.}~\bibnamefont {M\'elin}}, \bibinfo {author} {\bibfnamefont {A.~S.}\ \bibnamefont {Rashid}},\ and\ \bibinfo {author} {\bibfnamefont {M.}~\bibnamefont {Kayyalha}},\ }\bibfield  {title} {\bibinfo {title} {Ballistic andreev interferometers},\ }\href {https://doi.org/10.1103/PhysRevB.110.235419} {\bibfield  {journal} {\bibinfo  {journal} {Phys. Rev. B}\ }\textbf {\bibinfo {volume} {110}},\ \bibinfo {pages} {235419} (\bibinfo {year} {2024})}\BibitemShut {NoStop}%
\bibitem [{\citenamefont {Yamamoto}\ \emph {et~al.}(2012)\citenamefont {Yamamoto}, \citenamefont {Takada}, \citenamefont {B\"auerle}, \citenamefont {Watanabe}, \citenamefont {Wieck},\ and\ \citenamefont {Tarucha}}]{Yamamoto2012}%
  \BibitemOpen
  \bibfield  {author} {\bibinfo {author} {\bibfnamefont {M.}~\bibnamefont {Yamamoto}}, \bibinfo {author} {\bibfnamefont {S.}~\bibnamefont {Takada}}, \bibinfo {author} {\bibfnamefont {C.}~\bibnamefont {B\"auerle}}, \bibinfo {author} {\bibfnamefont {K.}~\bibnamefont {Watanabe}}, \bibinfo {author} {\bibfnamefont {A.~D.}\ \bibnamefont {Wieck}},\ and\ \bibinfo {author} {\bibfnamefont {S.}~\bibnamefont {Tarucha}},\ }\bibfield  {title} {\bibinfo {title} {{Electrical control of a solid-state flying qubit.}},\ }\href {https://doi.org/10.1038/nnano.2012.28} {\bibfield  {journal} {\bibinfo  {journal} {Nature Nanotech.}\ }\textbf {\bibinfo {volume} {7}},\ \bibinfo {pages} {247} (\bibinfo {year} {2012})}\BibitemShut {NoStop}%
\bibitem [{\citenamefont {Amado}\ \emph {et~al.}(2014)\citenamefont {Amado}, \citenamefont {Fornieri}, \citenamefont {Biasiol}, \citenamefont {Sorba},\ and\ \citenamefont {Giazotto}}]{Amado2014}%
  \BibitemOpen
  \bibfield  {author} {\bibinfo {author} {\bibfnamefont {M.}~\bibnamefont {Amado}}, \bibinfo {author} {\bibfnamefont {A.}~\bibnamefont {Fornieri}}, \bibinfo {author} {\bibfnamefont {G.}~\bibnamefont {Biasiol}}, \bibinfo {author} {\bibfnamefont {L.}~\bibnamefont {Sorba}},\ and\ \bibinfo {author} {\bibfnamefont {F.}~\bibnamefont {Giazotto}},\ }\bibfield  {title} {\bibinfo {title} {{A ballistic two-dimensional-electron-gas Andreev interferometer}},\ }\href {https://doi.org/10.1063/1.4884952} {\bibfield  {journal} {\bibinfo  {journal} {App. Phys. Lett.}\ }\textbf {\bibinfo {volume} {104}},\ \bibinfo {pages} {242604} (\bibinfo {year} {2014})}\BibitemShut {NoStop}%
\bibitem [{\citenamefont {Cheng}\ \emph {et~al.}(2009)\citenamefont {Cheng}, \citenamefont {Xing}, \citenamefont {Wang},\ and\ \citenamefont {Sun}}]{Cheng2009}%
  \BibitemOpen
  \bibfield  {author} {\bibinfo {author} {\bibfnamefont {S.-g.}\ \bibnamefont {Cheng}}, \bibinfo {author} {\bibfnamefont {Y.}~\bibnamefont {Xing}}, \bibinfo {author} {\bibfnamefont {J.}~\bibnamefont {Wang}},\ and\ \bibinfo {author} {\bibfnamefont {Q.-f.}\ \bibnamefont {Sun}},\ }\bibfield  {title} {\bibinfo {title} {{Controllable Andreev retroreflection and specular Andreev reflection in a four-terminal graphene-superconductor hybrid system}},\ }\href {https://doi.org/10.1103/PhysRevLett.103.167003} {\bibfield  {journal} {\bibinfo  {journal} {Phys. Rev. Lett.}\ }\textbf {\bibinfo {volume} {103}},\ \bibinfo {pages} {167003} (\bibinfo {year} {2009})}\BibitemShut {NoStop}%
\bibitem [{\citenamefont {Pandey}\ \emph {et~al.}(2019)\citenamefont {Pandey}, \citenamefont {Kraft}, \citenamefont {Krupke}, \citenamefont {Beckmann},\ and\ \citenamefont {Danneau}}]{Pandey2019}%
  \BibitemOpen
  \bibfield  {author} {\bibinfo {author} {\bibfnamefont {P.}~\bibnamefont {Pandey}}, \bibinfo {author} {\bibfnamefont {R.}~\bibnamefont {Kraft}}, \bibinfo {author} {\bibfnamefont {R.}~\bibnamefont {Krupke}}, \bibinfo {author} {\bibfnamefont {D.}~\bibnamefont {Beckmann}},\ and\ \bibinfo {author} {\bibfnamefont {R.}~\bibnamefont {Danneau}},\ }\bibfield  {title} {\bibinfo {title} {Andreev reflection in ballistic normal metal/graphene/superconductor junctions},\ }\href {https://doi.org/10.1103/PhysRevB.100.165416} {\bibfield  {journal} {\bibinfo  {journal} {Phys. Rev. B}\ }\textbf {\bibinfo {volume} {100}},\ \bibinfo {pages} {165416} (\bibinfo {year} {2019})}\BibitemShut {NoStop}%
\bibitem [{\citenamefont {Takagaki}(2023)}]{Takagaki2023}%
  \BibitemOpen
  \bibfield  {author} {\bibinfo {author} {\bibfnamefont {Y.}~\bibnamefont {Takagaki}},\ }\bibfield  {title} {\bibinfo {title} {Role of symmetry in quantum blocking of andreev reflection in graphene nanoribbons side-terminated by superconductors},\ }\href {https://doi.org/10.1088/1361-648X/accf59} {\bibfield  {journal} {\bibinfo  {journal} {J. Phys. Condens. Matter}\ }\textbf {\bibinfo {volume} {35}},\ \bibinfo {pages} {315301} (\bibinfo {year} {2023})}\BibitemShut {NoStop}%
\bibitem [{\citenamefont {Rashid}\ \emph {et~al.}(2024)\citenamefont {Rashid}, \citenamefont {Yi}, \citenamefont {Taniguchi}, \citenamefont {Watanabe}, \citenamefont {Samarth}, \citenamefont {Mélin},\ and\ \citenamefont {Kayyalha}}]{Rashid2024}%
  \BibitemOpen
  \bibfield  {author} {\bibinfo {author} {\bibfnamefont {A.~S.}\ \bibnamefont {Rashid}}, \bibinfo {author} {\bibfnamefont {L.}~\bibnamefont {Yi}}, \bibinfo {author} {\bibfnamefont {T.}~\bibnamefont {Taniguchi}}, \bibinfo {author} {\bibfnamefont {K.}~\bibnamefont {Watanabe}}, \bibinfo {author} {\bibfnamefont {N.}~\bibnamefont {Samarth}}, \bibinfo {author} {\bibfnamefont {R.}~\bibnamefont {Mélin}},\ and\ \bibinfo {author} {\bibfnamefont {M.}~\bibnamefont {Kayyalha}},\ }\href {https://arxiv.org/abs/2405.02975} {\bibinfo {title} {Exploring nonequilibrium andreev resonances in ultraclean graphene andreev interferometers}} (\bibinfo {year} {2024}),\ \Eprint {https://arxiv.org/abs/2405.02975} {arXiv:2405.02975 [cond-mat.mes-hall]} \BibitemShut {NoStop}%
\bibitem [{\citenamefont {Blasi}\ \emph {et~al.}(2020{\natexlab{a}})\citenamefont {Blasi}, \citenamefont {Taddei}, \citenamefont {Arrachea}, \citenamefont {Carrega},\ and\ \citenamefont {Braggio}}]{Blasi2020}%
  \BibitemOpen
  \bibfield  {author} {\bibinfo {author} {\bibfnamefont {G.}~\bibnamefont {Blasi}}, \bibinfo {author} {\bibfnamefont {F.}~\bibnamefont {Taddei}}, \bibinfo {author} {\bibfnamefont {L.}~\bibnamefont {Arrachea}}, \bibinfo {author} {\bibfnamefont {M.}~\bibnamefont {Carrega}},\ and\ \bibinfo {author} {\bibfnamefont {A.}~\bibnamefont {Braggio}},\ }\bibfield  {title} {\bibinfo {title} {Nonlocal thermoelectricity in a superconductor--topological-insulator--superconductor junction in contact with a normal-metal probe: Evidence for helical edge states},\ }\href {https://doi.org/10.1103/PhysRevLett.124.227701} {\bibfield  {journal} {\bibinfo  {journal} {Phys. Rev. Lett.}\ }\textbf {\bibinfo {volume} {124}},\ \bibinfo {pages} {227701} (\bibinfo {year} {2020}{\natexlab{a}})}\BibitemShut {NoStop}%
\bibitem [{\citenamefont {Blasi}\ \emph {et~al.}(2020{\natexlab{b}})\citenamefont {Blasi}, \citenamefont {Taddei}, \citenamefont {Arrachea}, \citenamefont {Carrega},\ and\ \citenamefont {Braggio}}]{Blasi2020b}%
  \BibitemOpen
  \bibfield  {author} {\bibinfo {author} {\bibfnamefont {G.}~\bibnamefont {Blasi}}, \bibinfo {author} {\bibfnamefont {F.}~\bibnamefont {Taddei}}, \bibinfo {author} {\bibfnamefont {L.}~\bibnamefont {Arrachea}}, \bibinfo {author} {\bibfnamefont {M.}~\bibnamefont {Carrega}},\ and\ \bibinfo {author} {\bibfnamefont {A.}~\bibnamefont {Braggio}},\ }\bibfield  {title} {\bibinfo {title} {Nonlocal thermoelectricity in a topological andreev interferometer},\ }\href {https://doi.org/10.1103/PhysRevB.102.241302} {\bibfield  {journal} {\bibinfo  {journal} {Phys. Rev. B}\ }\textbf {\bibinfo {volume} {102}},\ \bibinfo {pages} {241302} (\bibinfo {year} {2020}{\natexlab{b}})}\BibitemShut {NoStop}%
\bibitem [{\citenamefont {Plaszk\'o}\ \emph {et~al.}(2020)\citenamefont {Plaszk\'o}, \citenamefont {Rakyta}, \citenamefont {Cserti}, \citenamefont {Korm\'anyos},\ and\ \citenamefont {Lambert}}]{Plaszko2020}%
  \BibitemOpen
  \bibfield  {author} {\bibinfo {author} {\bibfnamefont {N.~L.}\ \bibnamefont {Plaszk\'o}}, \bibinfo {author} {\bibfnamefont {P.}~\bibnamefont {Rakyta}}, \bibinfo {author} {\bibfnamefont {J.}~\bibnamefont {Cserti}}, \bibinfo {author} {\bibfnamefont {A.}~\bibnamefont {Korm\'anyos}},\ and\ \bibinfo {author} {\bibfnamefont {C.~J.}\ \bibnamefont {Lambert}},\ }\bibfield  {title} {\bibinfo {title} {{Quantum interference and nonequilibrium Josephson currents in molecular Andreev interferometers}},\ }\href {https://doi.org/10.3390/nano10061033} {\bibfield  {journal} {\bibinfo  {journal} {Nanomaterials}\ }\textbf {\bibinfo {volume} {10}},\ \bibinfo {pages} {1033} (\bibinfo {year} {2020})}\BibitemShut {NoStop}%
\bibitem [{\citenamefont {Cioni}(2025)}]{myplottingtool2025}%
  \BibitemOpen
  \bibfield  {author} {\bibinfo {author} {\bibfnamefont {F.}~\bibnamefont {Cioni}},\ }\href {https://doi.org/10.5281/zenodo.15668452} {\bibinfo {title} {fcioni13/cooling-via-andreev-interferometer: First release (2025.06.15).}} (\bibinfo {year} {2025})\BibitemShut {NoStop}%
\bibitem [{\citenamefont {Lambert}\ and\ \citenamefont {Raimondi}(1998)}]{Lambert1998}%
  \BibitemOpen
  \bibfield  {author} {\bibinfo {author} {\bibfnamefont {C.~J.}\ \bibnamefont {Lambert}}\ and\ \bibinfo {author} {\bibfnamefont {R.}~\bibnamefont {Raimondi}},\ }\bibfield  {title} {\bibinfo {title} {Phase-coherent transport in hybrid superconducting nanostructures},\ }\href {https://doi.org/10.1088/0953-8984/10/5/003} {\bibfield  {journal} {\bibinfo  {journal} {J. Phys. Condens. Matter}\ }\textbf {\bibinfo {volume} {10}},\ \bibinfo {pages} {901} (\bibinfo {year} {1998})}\BibitemShut {NoStop}%
\end{thebibliography}%

\end{document}